\documentclass[twocolumn]{aastex631}

\usepackage{xcolor,tikz}
\usepackage{newtxtext,newtxmath}
\usepackage[T1]{fontenc}
\usepackage{gensymb}
\usepackage{tablefootnote}
\usepackage{makecell}
\usepackage{comment}
\usepackage{amsmath}
\usepackage{graphicx} 
\graphicspath{{./figures/}}

\begin{document}

\title{The Emerging Class of Double-Faced White Dwarfs}
\shorttitle{Double-Faced WDs}

\author[0000-0001-7143-0890]{Adam Moss}
\affiliation{Homer L. Dodge Department of Physics and Astronomy, University of Oklahoma, 440 W. Brooks St., Norman, OK 73019, USA}
\author[0000-0001-6098-2235]{Mukremin Kilic}
\affiliation{Homer L. Dodge Department of Physics and Astronomy, University of Oklahoma, 440 W. Brooks St., Norman, OK 73019, USA}
\author[0000-0003-2368-345X]{Pierre Bergeron}
\affiliation{Département de Physique, Université de Montréal, C.P. 6128, Succ. Centre-Ville, Montréal, Québec H3C 3J7, Canada}
\author[0009-0009-9105-7865]{Gracyn Jewett}
\affiliation{Homer L. Dodge Department of Physics and Astronomy, University of Oklahoma, 440 W. Brooks St., Norman, OK 73019, USA}
\author[0000-0002-4462-2341]{Warren R. Brown}
\affiliation{Center for Astrophysics | Harvard \& Smithsonian,
60 Garden Street, Cambridge, MA 02138, USA}
\shortauthors{Moss et al.}
\email{adam.g.moss-1@ou.edu}

\begin{abstract}
We report our findings on a spectroscopic survey of seven unresolved DA+DB binary white dwarf candidates. We have discovered extreme spectroscopic variations in one of these candidates, SDSS J084716.21+484220.40. Previous analysis failed to reproduce the optical spectrum using a single object with a homogeneous atmosphere. Our time-resolved spectroscopy reveals a double-faced white dwarf that switches between a DBA and DA spectral type over 6.5 or 8.9 hours due to varying surface abundances. We also provide time-series spectroscopy of the magnetic DBA, SDSS J085618.94+161103.6 (LB 8915), and confirm an inhomogeneous atmosphere. We employ an atmosphere model with hydrogen caps and a helium belt that yields excellent fits to our time-resolved spectra. We use the oblique rotator model to derive the system geometry for both targets. With the addition of these two objects, the emerging class of double-faced white dwarfs now consists of seven members. We summarize the properties of this new class of objects, and discuss how magnetism impacts the convective processes and leads to the formation of double-faced white dwarfs. We identify cooler versions of white dwarfs with inhomogeneous atmospheres among the cool magnetic DA white dwarf sample,
where the H$\alpha$ line is shallower than expected based on pure hydrogen atmosphere models. 
\end{abstract}
\date{July 2024}

\section{Introduction}
\label{evol}
The surface composition of a significant fraction of white dwarfs changes with time due to gravitational settling, radiative levitation, winds, convection, and external accretion. This leads to spectral evolution; a large fraction of white dwarfs transition from one spectral type to another. For example, a recent study of the 100 pc white dwarf sample in the SDSS footprint \citep{Kilic24} showed that $\approx70$\% of
white dwarfs retain hydrogen atmospheres throughout their evolution and 10\% retain helium atmospheres. In these objects, there is no mechanism that can alter the atmosphere composition  (besides trace amounts of metals that could be accreted from debris disks) given the thickness of their outer layers. Thus the dominant element in these atmospheres does not change at any point in the cooling sequence; these objects do not undergo spectral evolution. The remaining 20\% consists of stars where the aforementioned processes greatly alter the composition, resulting in a transition from one atmosphere type to another.

Due to the strong gravitational settling in white dwarfs, hydrogen gradually floats up to the outermost layers as a newly formed white dwarf cools. This float-up process helps DO white dwarfs with helium atmospheres transition into DA white dwarfs, provided there is enough hydrogen in the interior \citep{Fontaine87,Bedard24}. The location on the cooling sequence where this transition occurs depends on the initial hydrogen content, with higher mass fractions leading to transitions at hotter temperatures. For example, an initial mass fraction of $X_{\mathrm{H}} = 10^{-3}$ in the outer $10^{-4}~M_\star$ leads to a DO-DA transition at $T_{\rm eff}=70,000$ K, whereas $X_{\mathrm{H}} = 10^{-6}$ results in a transition much later, when the star cools down to $T_{\rm eff}=35,000$ K \citep{Bedard24}. 

While the float-up process can change a DO into a DA, the newly formed exterior hydrogen layer is relatively thin, and thus the target will likely undergo further spectral evolution as it continues to cool. If the hydrogen layer is thinner than the outer $\sim$$10^{-14}~M_\star$, the underlying helium convective zone will erode and dilute this layer later in the cooling sequence, resulting in a helium-dominated atmosphere (DB) or a mixed atmosphere (DBA). This convective dilution process is thought to occur between $30,000$ K $\gtrsim T_{\rm eff} \gtrsim 14,000$ K depending on the hydrogen layer mass \citep{Macdonald91,Rolland18,Rolland20,Bedard23}. 

Even if the hydrogen layer is massive enough to avoid convective dilution, DA white dwarfs with thin surface hydrogen layers have more chances to transition. In a hydrogen-rich white dwarf, the convection zone appears at about $T_{\rm eff} = 18,000$ K \citep{Cunningham19}, and it expands significantly below 12,000 K to include a large portion of the envelope. Hence, convective mixing can lead to further spectral evolution and increase the fraction of He-atmosphere white dwarfs at cooler temperatures \citep{Tremblay08,Chen11,Rolland18,Cunningham20,Bedard22,Bergeron22}.

Observational evidence for these processes manifests in how the ratio of spectral types varies across effective temperatures. \citet{Eisenstein06} found that the ratio of DAs to DBs is 2.5 times as high at 30,000 K compared to 20,000 K due to spectral evolution. \citet{Bedard20} analyzed 1806 white dwarfs at $T_{\rm eff} \geq 30,000$ K and found that $\sim$2/3 of DOs eventually become DAs as they cool due to the float-up process.  Multiple recent studies have further confirmed the steady increase of the fraction of DBs below 20,000 K \citep{Genest19,Ourique19,Cunningham20,Lopez22,Torres23,Obrien24,Vincent24}. While the details of how rapidly the fraction increases vary between studies depending on what sample is used, there is significant evidence for the transition of some DAs to DBs/DBAs due to convective dilution ($30,000$ K $\gtrsim T_{\rm eff} \gtrsim 14,000$ K) and convective mixing ($14,000$ K $\gtrsim T_{\rm eff} \gtrsim 6000$ K) .

Regardless of whether convective dilution or convective mixing occurs, the formation of mixed atmosphere white dwarfs is of particular interest for two reasons. One is that $60 - 75\%$ of the DB population are DBAs \citep{Koester15,Rolland18}, which means that the majority of helium-dominated objects contain enough hydrogen for the DO-DA transition to occur. The second reason is that recent modelling has been able to reproduce the observed hydrogen abundances using the aforementioned internal transport of hydrogen to the surface \citep{Rolland20,Bedard23}. By treating the dilution process with a deep hydrogen reservoir, the production of DBAs can be explained via internal processes as opposed to requiring external accretion \citep{Farihi13,Gentile17}.

These internal convective processes should lead to homogeneously mixed surface layers. Hence, recent discoveries of several white dwarfs with inhomogeneous atmospheres is surprising. These `double-faced' white dwarfs show evidence of varying H/He abundance ratios across the stellar surface in time-resolved spectra. The most extreme example is ZTF J203349.8+322901.1 \citep{Caiazzo23} in which the hydrogen and helium lines completely vanish and reappear, suggesting one side is comprised of hydrogen and the other of helium. Recent modelling suggests that a stratified atmosphere where the thickness of the surface hydrogen layer varies across the surface can explain all of the observations in this relatively hot white dwarf (A. B\'edard 2024, private communication). 

Other examples of double-faced white dwarfs include GD 323 \citep{Pereira05}, Feige 7 \citep{Achilleos92}, and GALEX J071816.4+373139 \citep{Cheng24}. The latter two are particularly noteworthy because they are magnetic, with field strengths of $B_{\rm d} = 35$ MG and $B_{\rm d} = 8$ MG, respectively. \citet{Achilleos92} attributed the inhomogeneous atmosphere of Feige 7 to magnetism, where the movement of helium from the interior is constrained along the field lines, resulting in a higher He abundance at the poles. Magnetism has also been invoked to explain spectral variations in two metal-polluted white dwarfs \citep{Bagnulo24a,Bagnulo24b}, with polar caps containing high metal abundances compared to the equatorial regions.

\citet{Moss24} recently discovered variations in the hydrogen line strength of the magnetic DBA white dwarf SDSS J091016.43+210554.20 with a field strength of $\sim$0.5 MG. This object is one of the 10 DA+DB white dwarf binary candidates from \citet{Genest19} where homogeneous atmosphere models fail to accurately reproduce the optical spectra under the assumption of a single star (see their Figure 25.4). \citet{Moss24} demonstrated that J0910+2105 is instead a double-faced white dwarf with a rotation period of 7.7 or 11.3 hours. The oblique rotator model \citep{Stibbs50,Monaghan73} with hydrogen polar caps and a helium equatorial belt provides excellent fits to their time-resolved spectra. 

Given that there are nine other unresolved DA+DB binary candidates in \citet{Genest19}, we initiated a time-resolved spectroscopy survey to understand their nature and the emerging class of double-faced white dwarfs. Specifically, as we find more of these double-faced objects, we can begin to understand the range of possible surface geometries (extent of the H and He caps/belts etc), their location on the cooling sequence, the impact of magnetism, and the presence or lack of photometric variability and its significance. In this paper we present our findings on six of the unresolved binary candidates listed in \citet{Genest19}, as well as the magnetic DBA white dwarf LB 8915. We detail our observations and target selection in Section 2, followed by the model atmosphere analysis in Section 3. In Section 4 we discuss our findings with respect to other variable mixed-atmosphere white dwarfs, and detail the properties of the emerging class of double-faced white dwarfs. We then conclude in Section 5. 

\begin{figure}
    \centering
    \includegraphics[width=3.4in, clip=true, trim=0.8in 0.1in 0.85in 0.75in]{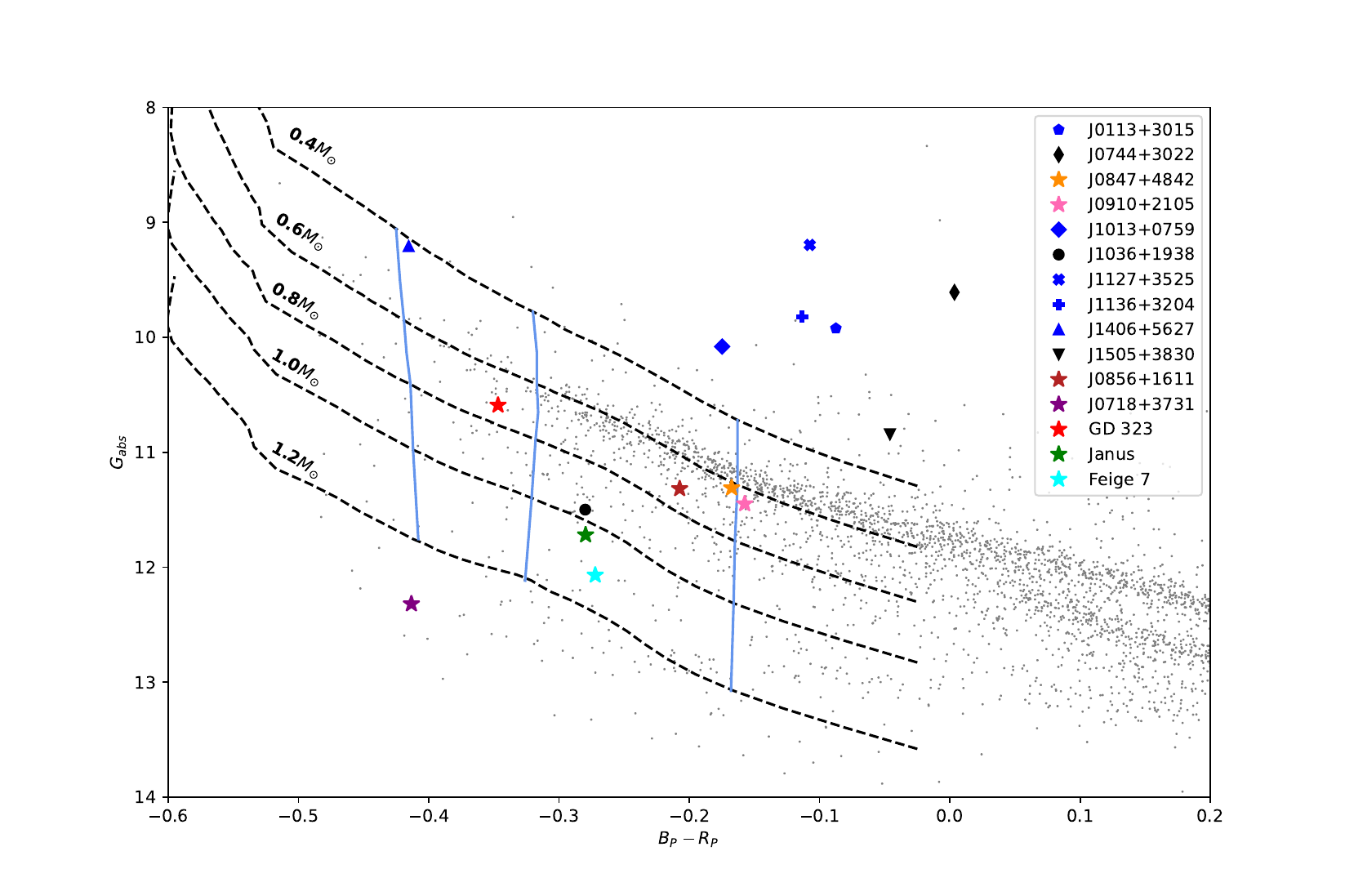}
    \caption{Gaia color-magnitude diagram of the unresolved DA+DB binary candidates from \citet{Genest19} along with the 100 pc white dwarf sample (gray points). Blue symbols mark targets that do not show variations in our time-series spectroscopy, whereas the black symbols are the remaining candidates with no follow-up data. Star symbols are the now seven confirmed double-faced white dwarfs, including J0847+4824 and J0856+1611 presented in this work. The evolutionary models for 0.4, 0.6, 0.8, 1.0, and 1.2 $M_{\odot}$ white dwarfs with pure He atmospheres are also shown. We plot the models down to 11,000 K, below which He lines disappear. The blue lines are isotherms assuming He atmospheres for 35,000, 25,000, and 15,000 K (left to right).}
    \label{fig1} 
\end{figure}

\section{Observations and Target Selection}

Figure \ref{fig1} shows our target selection in the Gaia color-magnitude diagram, along with several inhomogeneous atmosphere white dwarfs with confirmed spectral variations in the literature. We also show isotherms and evolutionary models for various masses assuming pure He atmospheres\footnote{See http://www.astro.umontreal.ca/$\sim$bergeron/CoolingModels.} \citep{Bergeron11,Bedard20}. Seven of the candidates from \citet[][black and blue symbols]{Genest19} are in fact overluminous; we have time-series spectra for five of these and none show spectroscopic variations over a few hours. We show in Section 3 that these five targets are likely unresolved binaries. Two of the targets however, J0847+4824 and J0856+1611, lie close together in the color-magnitude diagram, along with the double-faced white dwarf J0910+2105 \citep{Moss24}, and we show below that they are also double-faced stars. Among the remaining three candidates from \citet{Genest19} that we could not observe, one of them (J1036+1938) appears to be located close to ZTF J203349.8+322901.1 in the color-magnitude diagram, but follow-up time-series spectroscopy is required to confirm its nature. 

\begin{table*}%[!hp]
\begin{center}
\caption{\label{tab1} Observational Details} 
\begin{tabular}{|c|c|c|c|c|c|c|} 
 \hline
Object & Gaia DR3 ID & RA(J2000) & Dec(J2000) & APO & Gemini & MMT\\
 \hline
 J0113+3015 & 310054341933521664 & 01:13:56.39 & +30 15 14.63 & \makecell[l]{$11 \times 10$ min UT20231108 \\ $3\times 15$ min UT20231117} & $-$ &  $-$ \\ 
 \hline
 J0847+4842 & 1015028491488955776 & 08:47:16.18 & +48 42 20.43 & \makecell[tl]{$9 \times 15$ min UT20240305 \\ $4 \times 15$ min UT20240404 \\ $8 \times 15$ min UT20240405 \\ $16 \times 15$ min UT20240417} & $46 \times 3$ min & \makecell[tl]{12 exp \\ UT20240501 \\ $-$ UT20240506}\\
 \hline
 J0856+1611 & 611401999180118528 & 08:56:18.91 & +16:11:03.25 & \makecell[l]{$12\times10$min UT20240305 \\ $3\times10$min UT20240404 \\ $7\times10$min UT20240405} & $-$ & \makecell[l]{10 exp \\ UT20240501 \\ $-$ UT20240505} \\ 
 \hline
 J1013+0759 & 3874193610618742400 & 10:13:16.00 & +07:59:15.03 & $-$ & $47\times3$ min & $-$ \\
 \hline
 J1127+3252 & 4025468887134211840 & 11:27:11.70 & +32:52:29.45 & $-$&$8\times5$ min & $-$\\
 \hline
 J1136+3204 & 4024472523440858112 & 11:36:23.53 & +32:04:03.78 & $-$ &$23\times3$ min & $-$ \\ 
 \hline
 J1406+5627 & 1657929398564495872 & 14:06:15.81 & +56:27:25.87 & $-$&$35\times4$ min & $-$ \\
 \hline
\end{tabular}
\end{center}
\end{table*}

Table \ref{tab1} shows our observing summary for each target. We obtained multiple sequences for three targets at the Apache Point Observatory (APO) 3.5m telescope equipped with the Kitt Peak Ohio State Multi-Object Spectrograph (KOSMOS) over multiple nights. We used the 2" slit in the Center position with the Blue disperser and 2x2 binning. The setup covers the wavelength range 3800 - 6600 {\AA} with a resolution of 4.7 {\AA}.

We also obtained additional exposures for two targets at the 6.5m MMT with the Blue Channel Spectrograph. We used the 1.25" slit with the 500 l mm$^{-1}$ grating which yields a resolution of 4.5 {\AA}.

Finally, we obtained back-to-back sequences for five targets using the 8m Gemini North telescope with the Gemini Multi-Object Spectograph (GMOS) as part of the programs GN-2024A-Q-229, GN-2024A-Q-325, and GS-2024A-Q-329. We used the B480 grating with the 1" slit and 4x4 binning, providing wavelength coverage from 3555 - 7300 {\AA} and a resolution of 6.3 {\AA}. All of our spectra were reduced using standard IRAF procedures.

\section{Analysis}

For our variable targets, we start by fitting a Gaussian profile to both H$\beta$ and \ion{He}{1} $\lambda$4922 using LMFIT \citep{newville14} and calculate the equivalent width of these profiles for each spectrum. We then take the ratio of the equivalent widths to measure how the strength of the H lines vary as a function of time compared to the He lines. To calculate errors, we use bootstrapping to construct a distribution of values. We randomly sample the wavelength-flux pairs of a given spectrum, with replacement, while keeping only the unique pairs. This creates a new spectrum which we then fit as mentioned, repeating this process 10,000 times to generate 10,000 measurements. We then select the values at 15.9 and 84.1\% in the distribution as the $1\sigma$ errors. We then generate a Lomb-Scargle periodogram with the orbital fit code MPRVFIT \citep{DeLee13} to constrain the rotation period using the measured equivalent widths. 

Given that homogeneous atmosphere models failed to reproduce the optical spectra in \citet{Genest19} for these objects, we employ a polar cap model to generate fits to our time-resolved spectra, similar to \citet{Moss24}. In this model, the polar cap is made up entirely of H and the equatorial belt of He. We denote $\theta_{\rm c}$ as the extent of the polar cap from the pole down to the equator, such that $\theta_{\rm c} = 90\degree$ means the entire hemisphere is H. We also denote $\alpha$ as the angle between the magnetic axis and the plane of the sky, as defined in \citet[][$\alpha=90\degree$ means pole-on]{Schnerr06}. The purpose of using this geometry is to provide a physical mechanism for the observed variations. While the total amount of H and He is what determines the strength of each absorption line in our synthetic spectra, this by itself does not explain why we see varying abundances as the objects rotate. The specific use of H polar caps, positioned with respect to the magnetic axis which is misaligned with the rotation axis, accomplishes three goals: it provides excellent fits to our spectra, it explains why we see varying abundances as opposed to a constant ratio, and it fits within our framework of magnetically-inhibited convection at the poles (see Section \ref{4.2}).

We use the photometric technique described in \citet{Bergeron19} to obtain precise $T_{\rm eff}$ and $\log{g}$ values for the variable targets. We use the Gaia DR3 parallaxes, SDSS $u$, and Pan-STARRS $grizy$ photometry to constrain the effective temperature and solid angle of each object. Since the distance is known, we directly constrain the radius and then calculate the mass using white dwarf evolutionary models.

To generate the spectroscopic fits, the emergent Eddington flux $H_\nu$ is calculated by numerically integrating the specific intensity $I_\nu$ over the visible surface of the disk given the geometry and viewing angle. We adopt pure H and pure He model atmospheres based on the surface composition of each given element, and fix the $T_{\rm eff}$ and $\log{g}$ of both sets of models to the values obtained from the photometric fits. The details of our model grid are discussed further in \citet{Bergeron19} and \citet{Genest19}. This approach was successful in reproducing the time-resolved spectra for J0910+2105 \citep{Moss24}. 

\begin{figure*}
 \centering
 \includegraphics[width=3.3in, clip=true, trim=0.2in 0in 0.4in 0.5in]{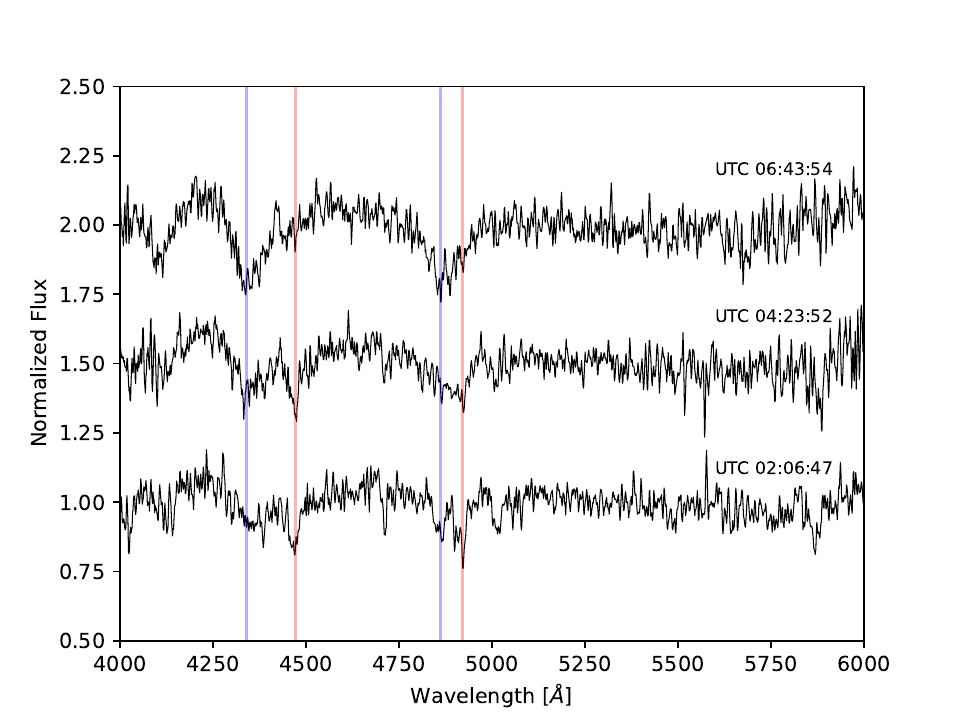}
 \includegraphics[width=3.75in, clip=true, trim=0.2in 0in 0.7in 0.55in]{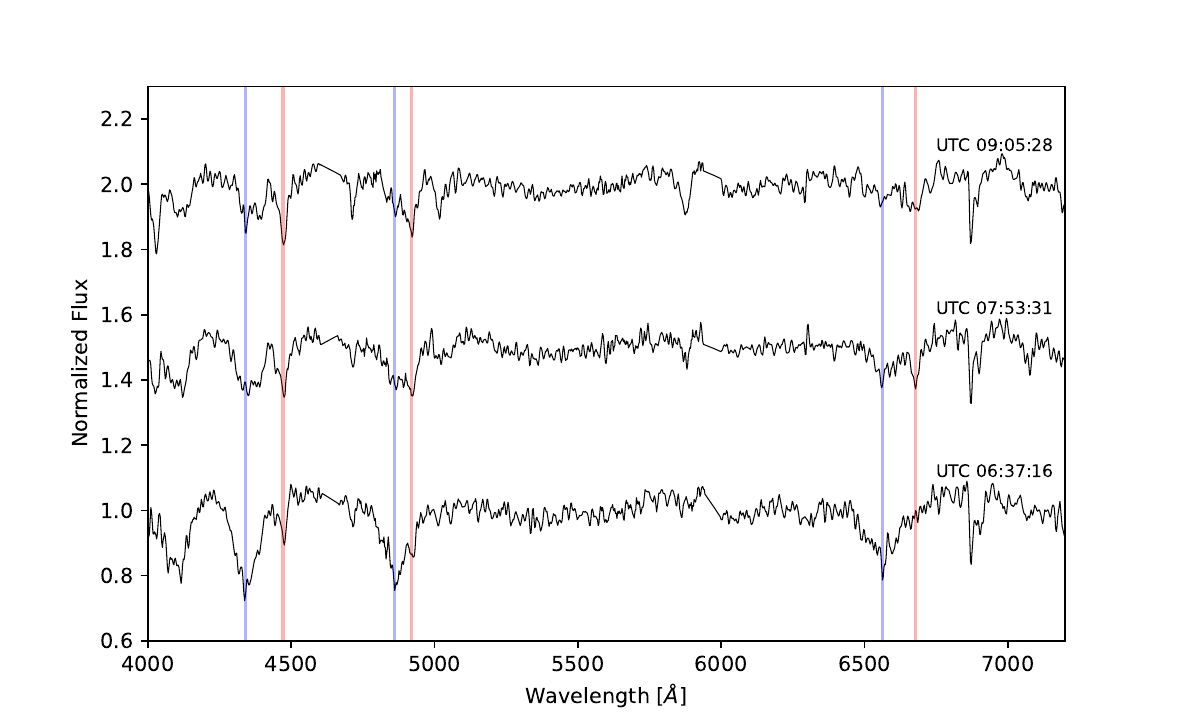}
 \includegraphics[width=3.5in, clip=true, trim=0.2in 0in 0in 0.55in]{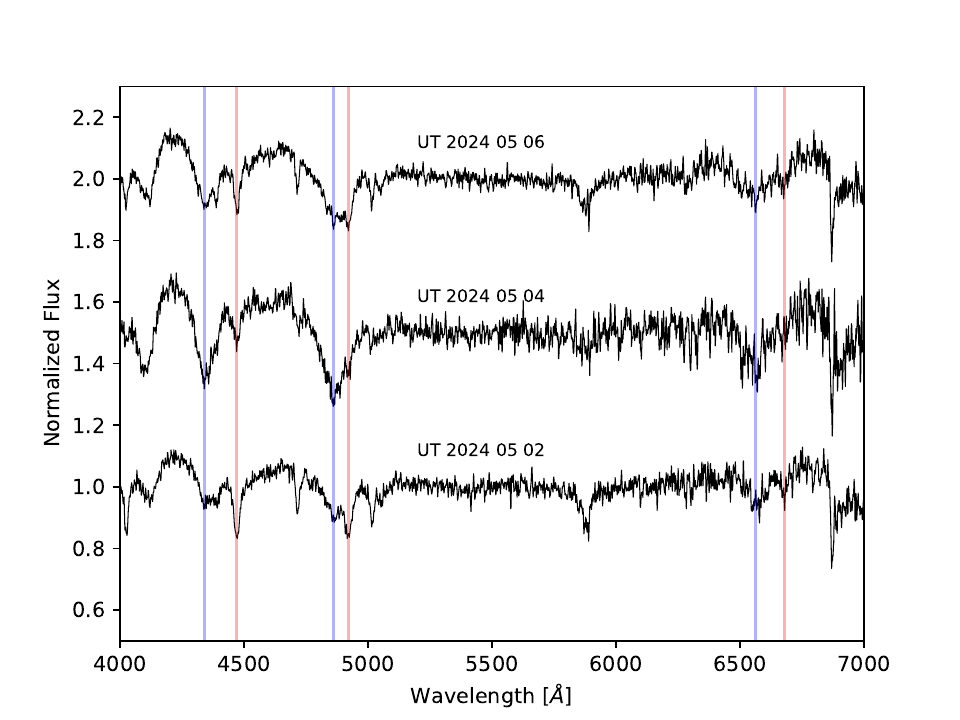}
 \caption{J0847+4842 sample spectra from the beginning, middle, and end of the night on UT 2024 March 05 (top left panel, APO) and 2024 February 18 (top right panel, Gemini). Sample spectra from separate nights at the MMT are shown in the bottom panel. The spectra are offset and smoothed for display purposes, and are plotted chronologically from bottom to top. The positions of the Balmer lines are marked with blue lines, and \ion{He}{1} $\lambda$4471, $\lambda$4922, and $\lambda$6678 are marked with red lines.}
 \label{fig2}
\end{figure*}

\begin{figure*}
\hspace{-0.2in}
 \includegraphics[width=2.5in, clip=true, trim=0.2in 0in 0.4in 0.5in]{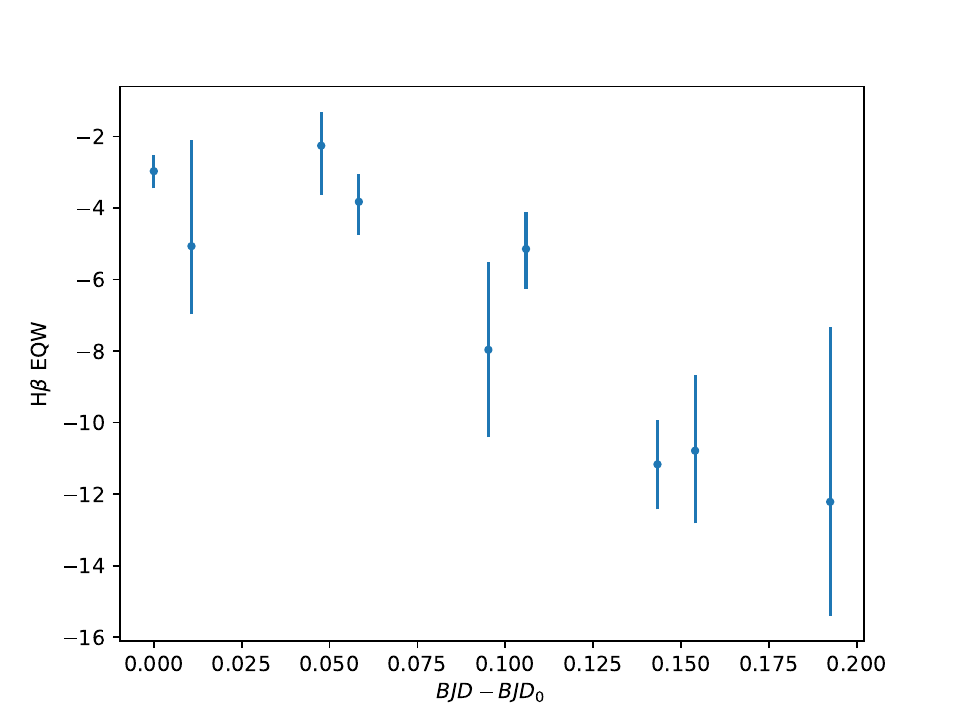}
 \includegraphics[width=2.5in, clip=true, trim=0.2in 0in 0.4in 0.5in]{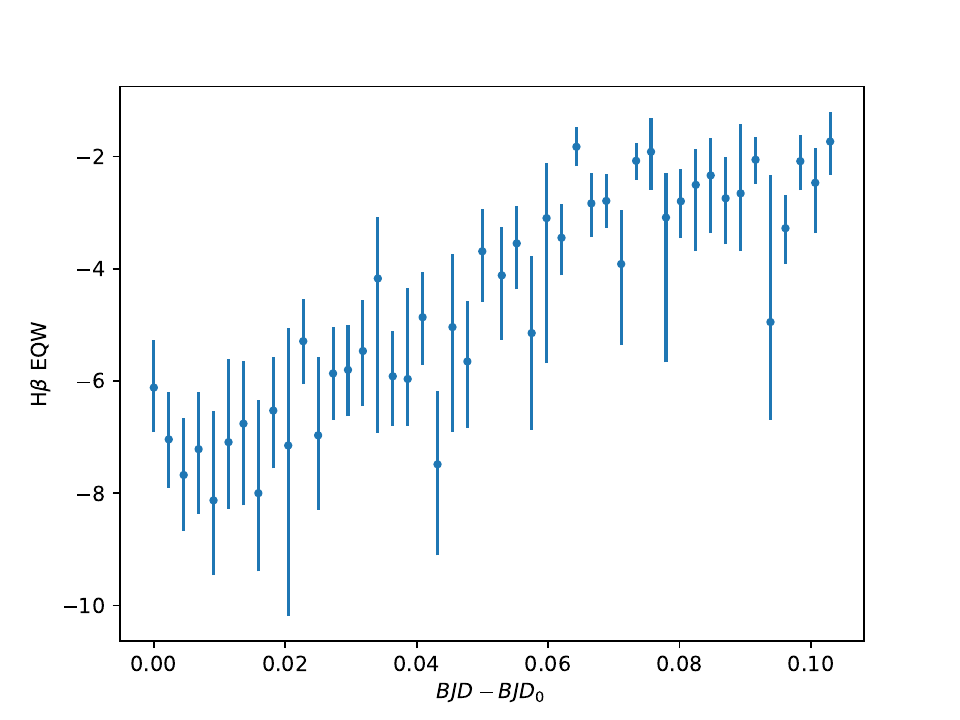}
  \includegraphics[width=2.5in, clip=true, trim=0.8in 0.6in 0.8in 1.1in]{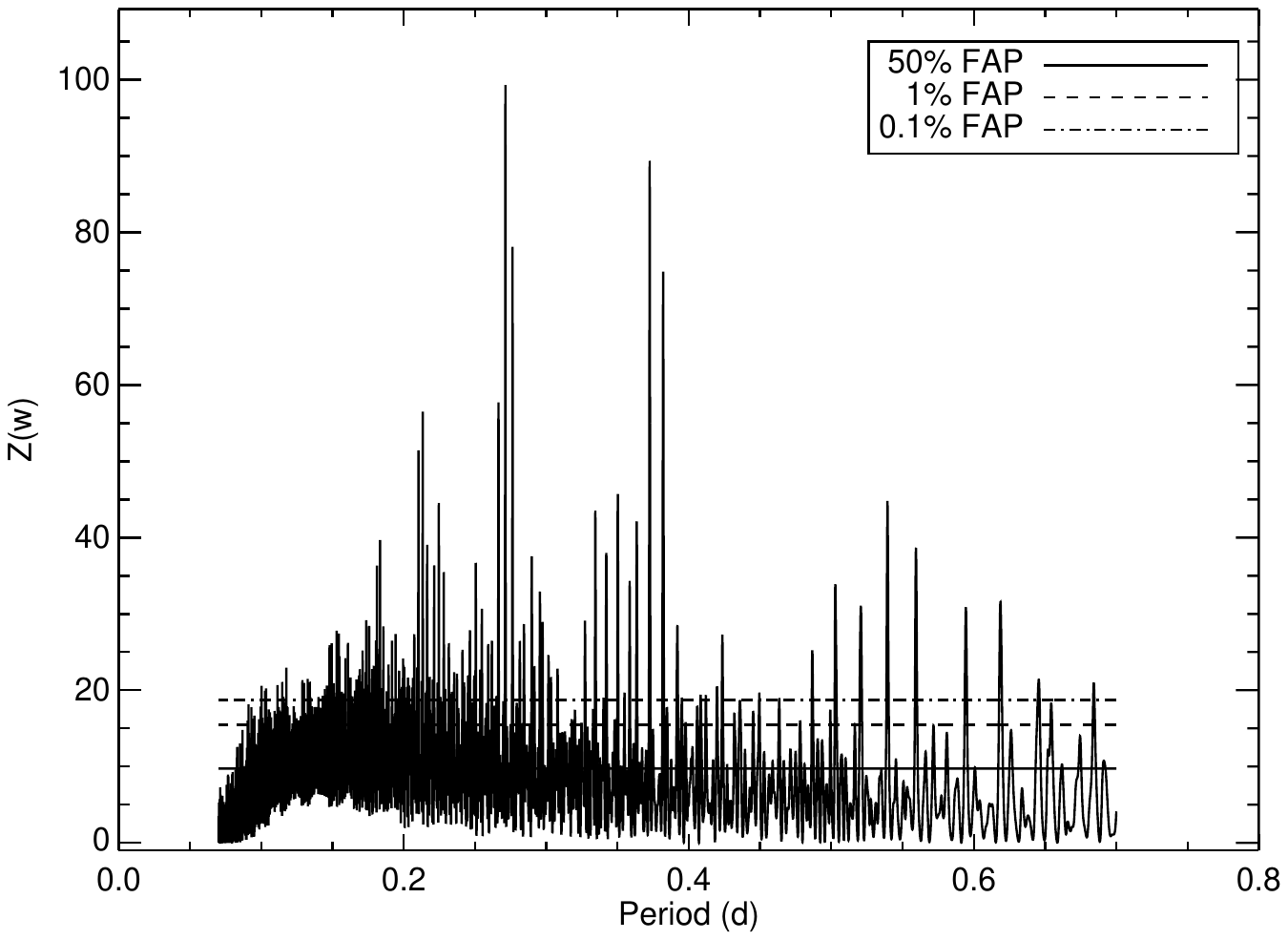}
 \caption{Equivalent width of the H$\beta$ line for J0847+4842 from UT 2024 March 05 (left) and February 18 (middle). The right panel shows the Lomb-Scargle periodogram including all of the APO, Gemini, and MMT data on this target. Despite aliasing, current data favor a rotation period of either 6.5 or 8.9 h.}
 \label{fig3}
\end{figure*}

With the $\alpha$ values found in our model fits, we then use the oblique rotator model \citep{Stibbs50,Monaghan73} to constrain the orientation of the line of sight and magnetic axis in each target. Variations in the spectra occur if the magnetic axis is offset from the rotation axis, so this model is key for providing a physical justification of any variations we observe. 

It is important to note two distinctions here. The first is that due to the degeneracies in the model fits (H caps versus He belts, and vice versa), we only use the H polar cap model. While our model only serves as a proxy for the possible geometries, this model provides excellent fits to our data. 

Second, \citet{Moss24} tested both convective and radiative atmospheres for J0910+2105, which has a magnetic field strength of $\sim$0.5 MG. While
radiative atmospheres provided a noticeable improvement in the fit to the \ion{He}{1} $\lambda$6678 line, the improvement with a radiative atmosphere is negligible outside of this line \citep[see also the discussion in][]{Lecavalier17}. In addition, our model fits are limited to wavelengths shorter than 5150 {\AA}. Our APO data only goes to 6600 {\AA}, which barely covers H$\alpha$ and does not include the \ion{He}{1} $\lambda$6678 line. To perform a uniform analysis of all of our APO, Gemini, and MMT data, we restrict our model fits to the wavelengths below 5150 {\AA}, where the majority of the He and H lines are. Hence we use convective atmosphere models in our fits. Fits for our variable targets, as well as all observed data are provided on Zenodo via the DOI \url{https://doi.org/10.5281/zenodo.14232043}.

\subsection{SDSS J084716.21+484220.40}

J0847+4842 is the most interesting target in our sample. Figure \ref{fig2} shows sample spectra from the beginning, middle, and end of the night on UT 2024 March 05 (left panel, APO) and 2024 February 18 (right panel, Gemini). We see both H and He lines at the start of the sequence shown in the left panel. As the night progresses, the He lines become much weaker, to the point where the spectrum looks like a DA at the end of the sequence. These variations are clearly due to the rotation of an inhomogeneous mixed atmosphere, and not an unresolved DA+DB binary. We captured the second half of the rotation phase in the Gemini data shown in the right panel. Here the sequence begins with strong H lines, which gradually become weaker while the He lines get stronger. 

Figure \ref{fig3} shows the equivalent width measurements from two different nights, along with the Lomb-Scargle diagram (right panel) including all of the data on J0847+4842. Because the \ion{He}{1} $\lambda$4922 line becomes nearly invisible, here we only use the H$\beta$ line. There is a clear increase in the H$\beta$ line strength (more negative equivalent width) with time in the left panel and a decrease in the right panel. The Lomb-Scargle diagram shown in the right panel reveals two significant peaks at 0.271 days (6.5 hours) and 0.373 days (8.9 hours). 

\begin{figure*}[!ht]
 \includegraphics[width=2.4in, clip=true, trim=1.1in 3.4in 0.9in 3.4in]{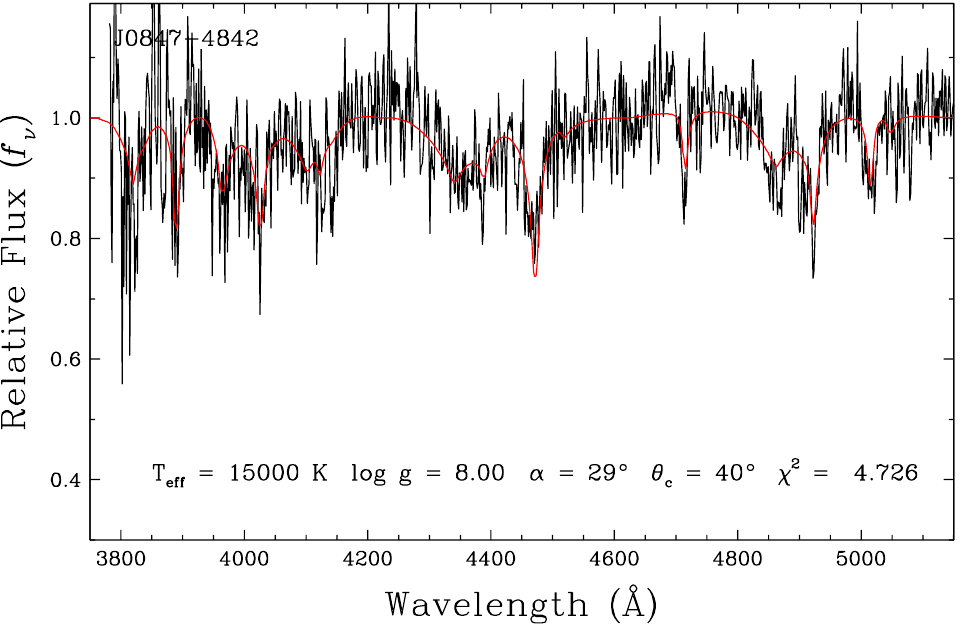}
 \includegraphics[width=2.4in, clip=true, trim=1.1in 3.4in 0.9in 3.4in]{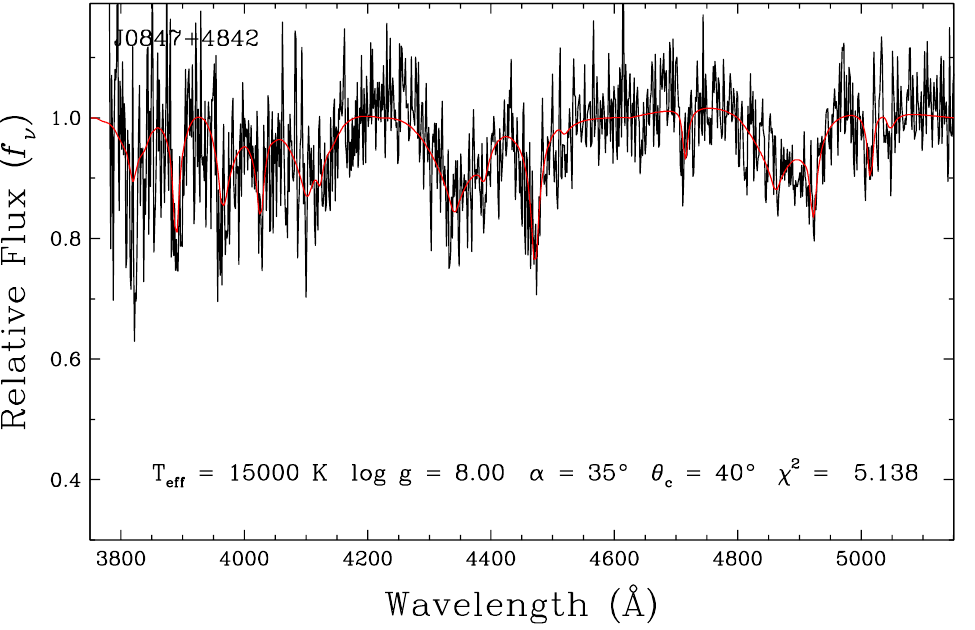}
 \includegraphics[width=2.4in, clip=true, trim=1.1in 3.4in 0.9in 3.4in]{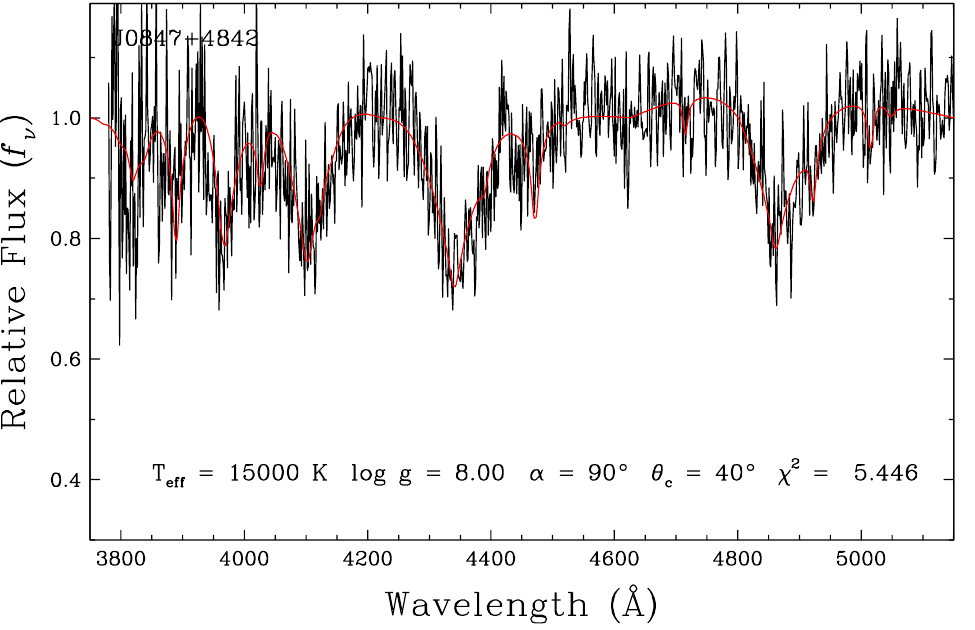}
 \includegraphics[width=2.4in, clip=true, trim=1.1in 3.4in 0.9in 3.4in]{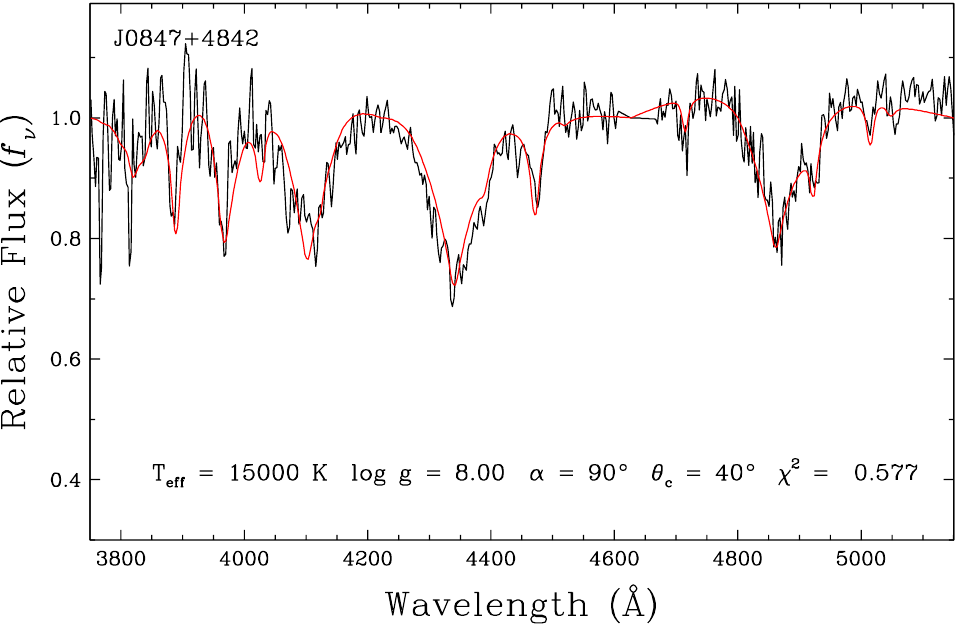}
 \includegraphics[width=2.4in, clip=true, trim=1.1in 3.4in 0.9in 3.4in]{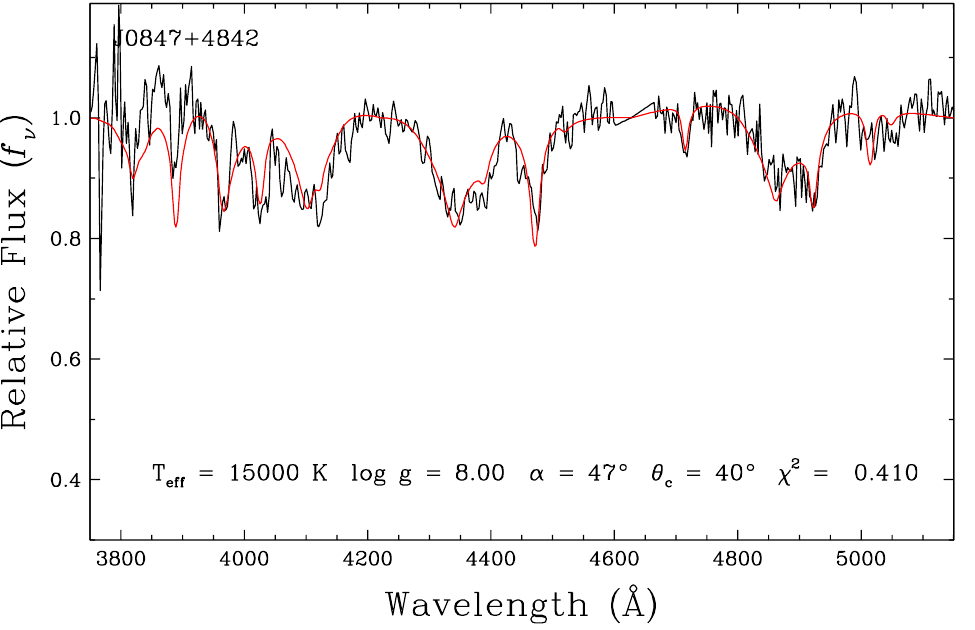}
 \includegraphics[width=2.4in, clip=true, trim=1.1in 3.4in 0.9in 3.4in]{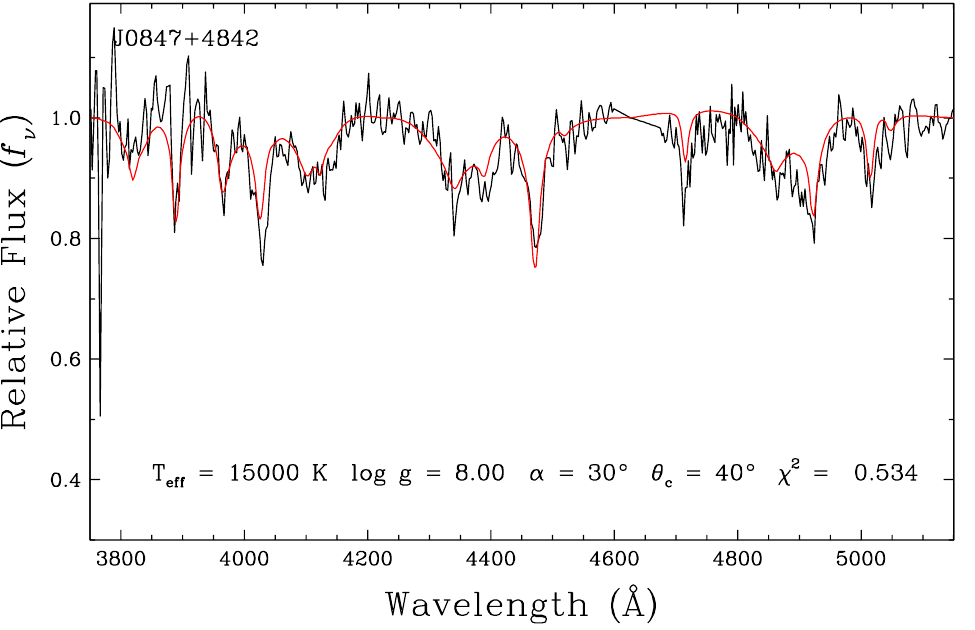}
 \caption{Example fits for J0847+4842 spectra from APO (top) and Gemini (bottom panels). Spectra are chronologically ordered from left to right. We fix the cap size to $\theta_{\rm c} = 40\degree$ and allow $\alpha$ to vary. An $\alpha$ angle near $90\degree$ means we are observing towards the pole, which results in stronger H lines in the spectra in our H polar cap model.}
 \label{fig4}
\end{figure*}

The photometric fit using the Gaia DR3 distance, SDSS $u$, and Pan-STARRS $grizy$ photometry indicate a relatively cool, average mass DBA white dwarf with ($T_{\rm eff} = 15,000$ K, $\log{g}=$ 7.99, and $M=0.585\ M_\odot$). We assume $\log{\rm H/He} = - 3$ for this fit. We use these parameters to generate our spectral fits using the polar cap model. Figure \ref{fig4} shows our fits to the APO and Gemini spectra shown in Figure \ref{fig2}. Unlike the homogeneous H/He atmosphere fit shown in \citet{Genest19}, we obtain excellent fits using our H polar cap model. We fix the cap size to $\theta_{\rm c} = 40\degree$ since this cap size best fits our entire data set and then allow $\alpha$ to vary.

Several spectra show strong H lines and nearly undetectable He lines, this implies that we see only the H polar caps in some rotation phases, while the He equatorial region is hidden. We obtain $\alpha \approx 90\degree$ for these spectra. While in other phases, the He lines are stronger, indicating an $\alpha$ value as low as $24^{\circ}$. We use the oblique rotator model to determine the geometry of this system, in which the magnetic axis is tilted with respect to the rotation axis. The model denotes $\beta$ as the angle between the two axes, and $i$ as the angle between the line of sight and rotation axis (see Figure 2 of \citealt{Bailey11} for an image of the geometry of this model). As the object rotates, the angle between the magnetic axis and the line-of-sight ranges from $\beta - i$ to $\beta + i$. This angle is equal to $90 - \alpha$, hence the range of $\alpha$ values from our fits constrains the system geometry. The minimum and maximum $\alpha$ values of 24\degree\ and 90\degree\ result in $\beta = 33\degree$ and $i = 33\degree$. This means the magnetic pole and the line of sight angle are aligned, and at certain rotation phases we are looking directly at the magnetic pole. This explains why we see almost exclusively H in some spectra.

\subsection{SDSS J085618.94+161103.6} 

J0856+1611 (LB 8915, erroneously referred to as LB 8827 in the literature) is listed as a DBA white dwarf in \citet{Genest19}, but it was not identified as a DA+DB binary candidate. Even though the photometric and spectroscopic temperatures agree within the errors, the mass estimates, 0.805 versus $1.060~M_\odot$ differ significantly.
This is likely due to the complications from the changing strengths of the H and He lines in this system. \citet{Putney97} showed this
object to be magnetic, and \citet{Wesemael01} demonstrated that Balmer line strengths vary with time, though this was based on only
three spectra. They concluded that these changes must be due to changes in the apparent H/He abundance ratio or irregularities over the stellar surface.

\begin{figure*}
 \includegraphics[width=2.5in, clip=true, trim=0.3in 0in 0.4in 0.5in]{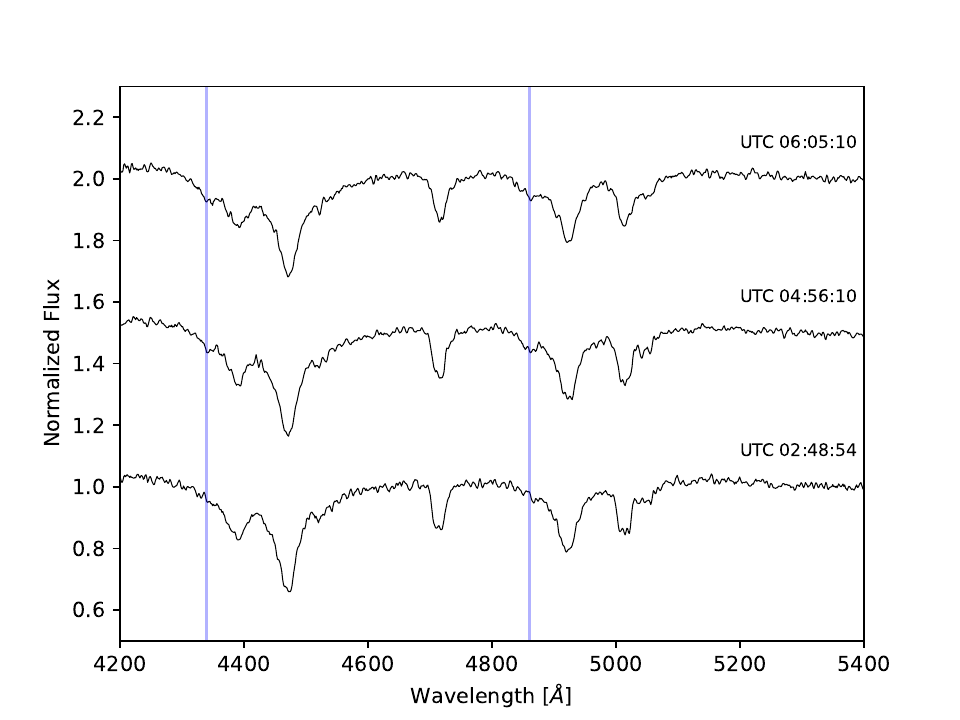}
 \includegraphics[width=2.5in, clip=true, trim=0.2in 0in 0.4in 0.5in]{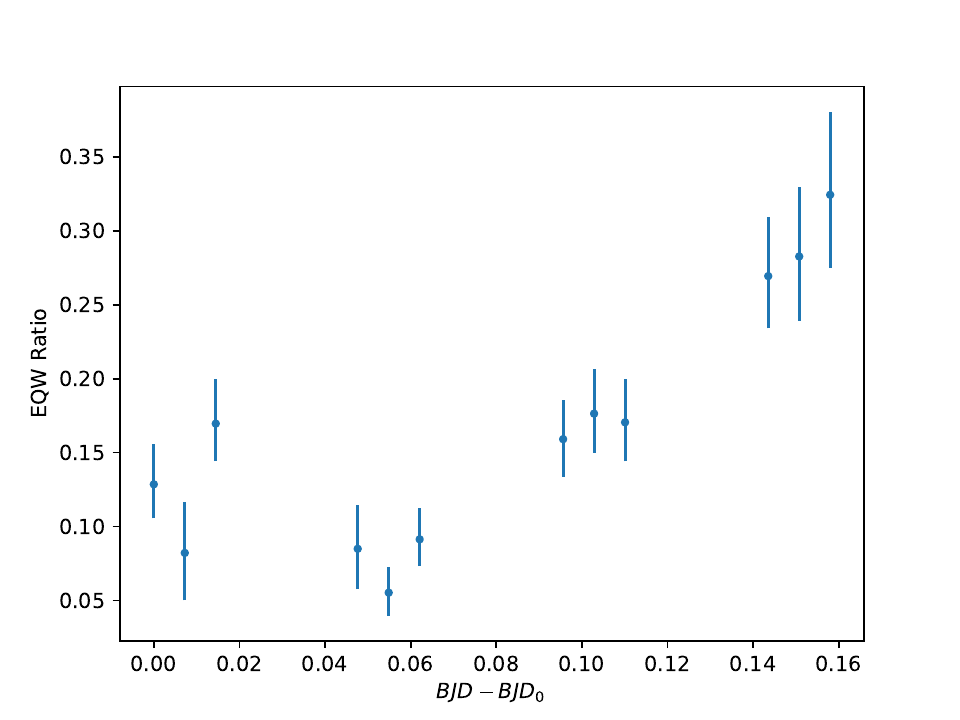}
 \includegraphics[width=2.5in, clip=true, trim=0.8in 0.6in 0.8in 1.1in]{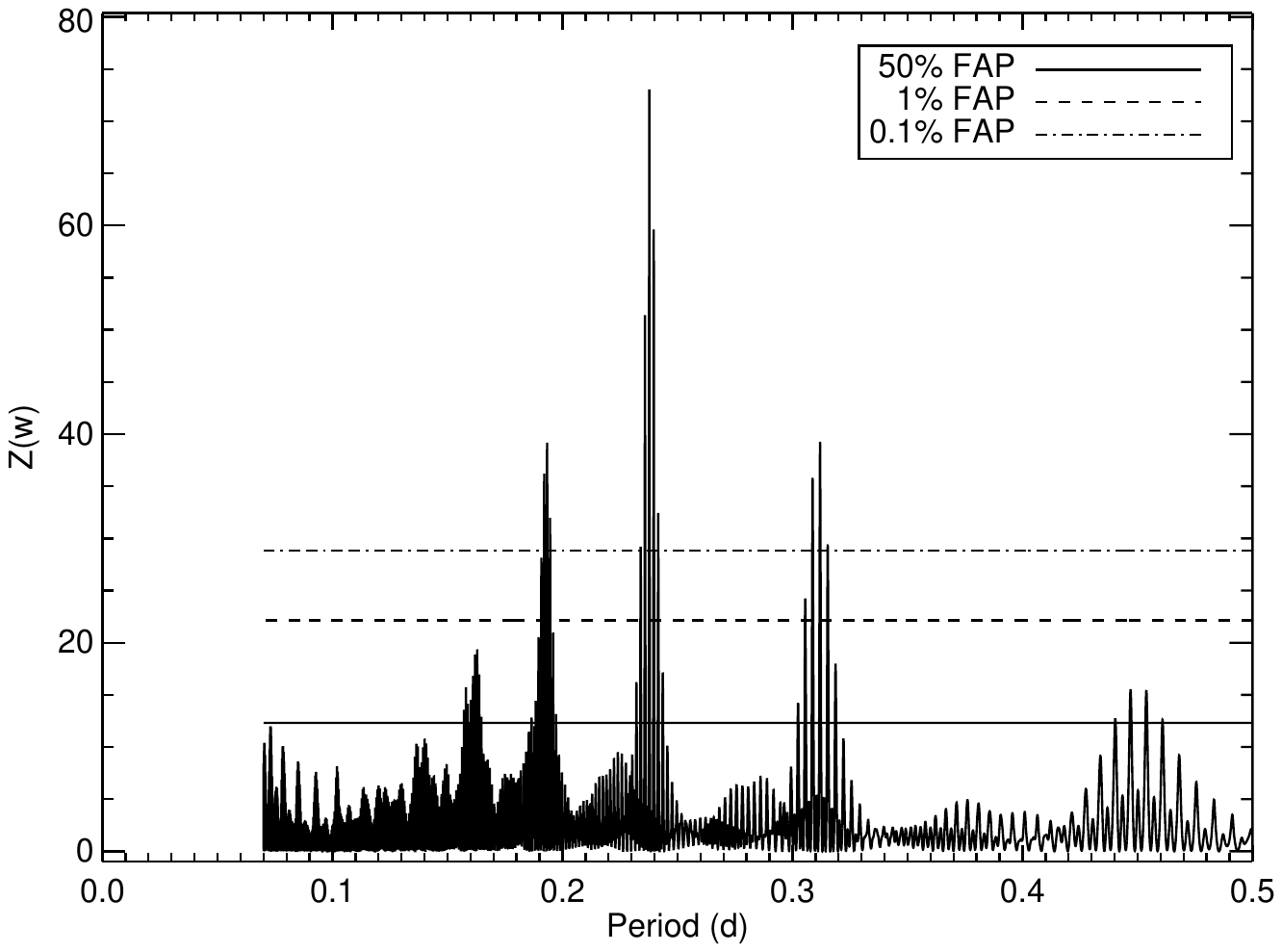}
 \caption{{\it Left:} J0856+1611 spectra from the beginning, middle, and end (bottom to top) of the APO sequence on UT 2024 March 05. Blue lines mark H$\gamma$ and H$\beta$. {\it Middle:} Equivalent width ratios for the UT 2024 March 05 sequence. {\it Right:} Lomb-Scargle periodogram. We successfully retrieve the 5.7 hour rotation period \citep{Hermes17a} from our entire data set.}
 \label{fig7}
\end{figure*}

To significantly improve the physical constraints on this system, we obtained 32 follow-up spectra over several nights. Figure \ref{fig7} shows three of the spectra from a sequence taken at APO on UT 2024 March 05. H lines are relatively weak in all of the spectra, however there are noticeable changes in the line strength throughout the course of the observations. H$\gamma$ in particular does not appear at all in several exposures, such as at the start of the sequence shown in this figure, but it is clearly visible later on, whereas He lines stay consistently strong. The middle panel in Figure \ref{fig7} shows the equivalent width ratios of H$\beta$ to \ion{He}{1} $\lambda$4922 from UT 2024 March 05, and the Lomb-Scargle periodogram generated from all of our data on this object. There is a clear increase in the H content as the object rotates, and we recover the 5.7 hour period as measured from K2 photometry \citep[see their Figure 3, labeled PG 0853+164]{Hermes17a}.  

\begin{figure*}%[h!]
\includegraphics[width=2.4in, clip=true, trim=1.1in 3.4in 0.9in 3.4in]{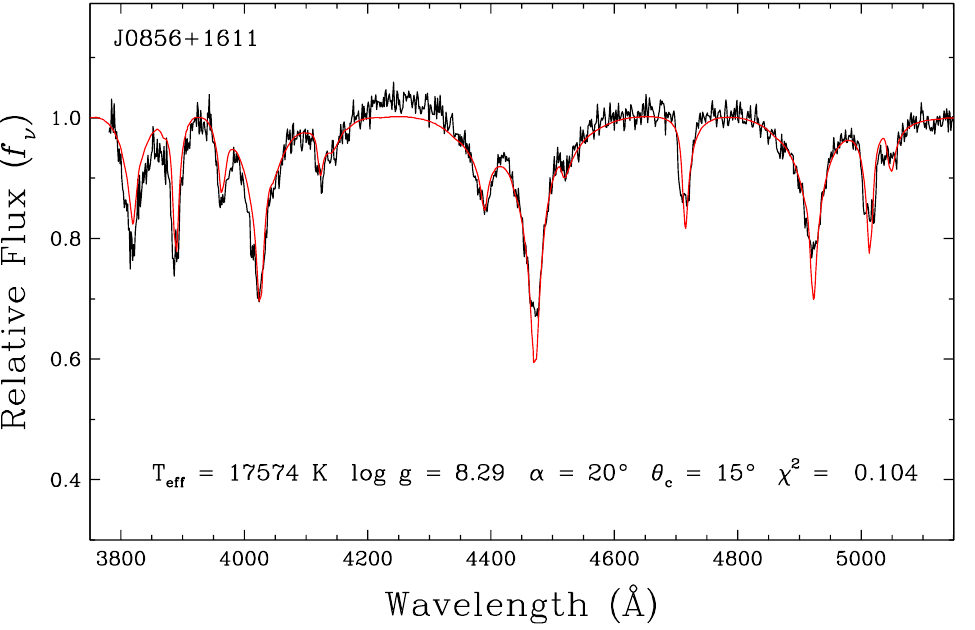}
\includegraphics[width=2.4in, clip=true, trim=1.1in 3.4in 0.9in 3.4in]{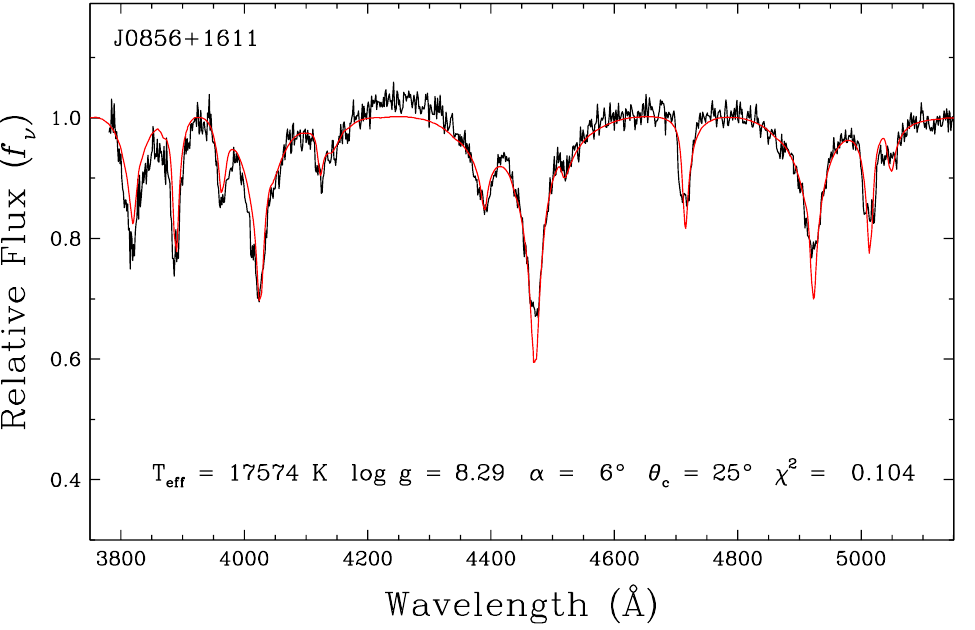}
\includegraphics[width=2.4in, clip=true, trim=1.1in 3.4in 0.9in 3.4in]{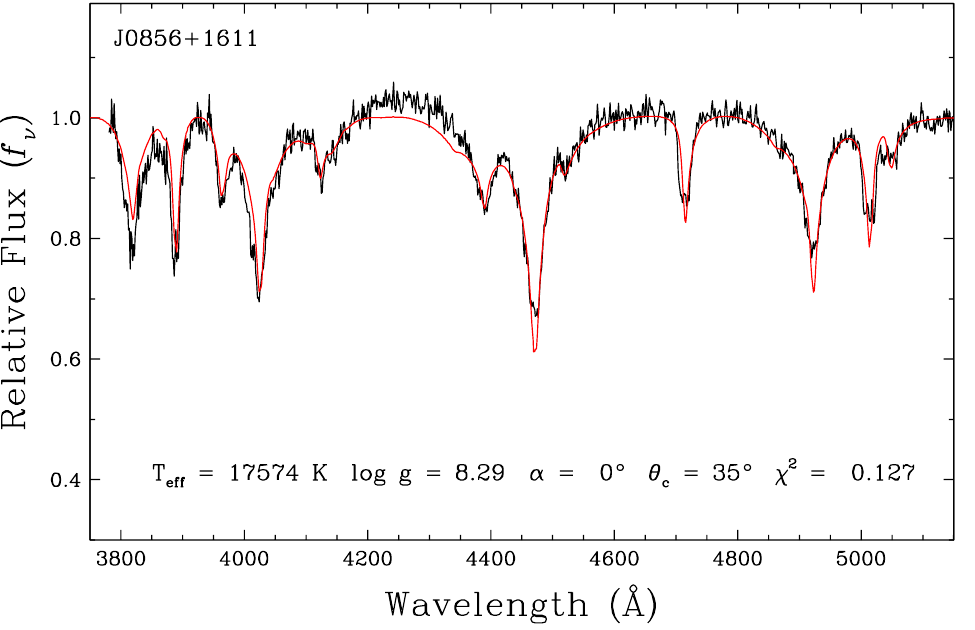}
\includegraphics[width=2.4in, clip=true, trim=1.1in 3.4in 0.9in 3.4in]{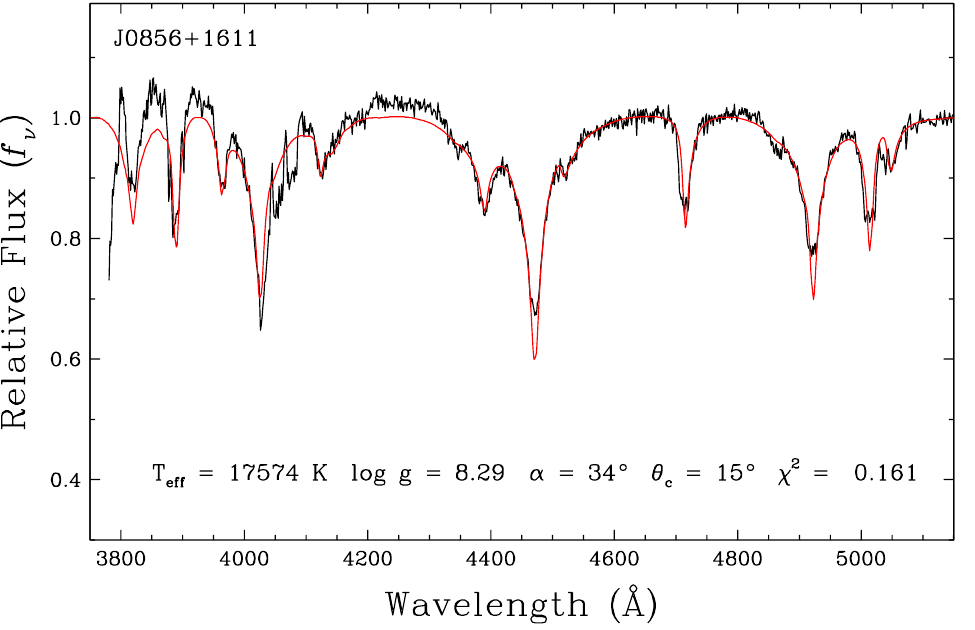}
\includegraphics[width=2.4in, clip=true, trim=1.1in 3.4in 0.9in 3.4in]{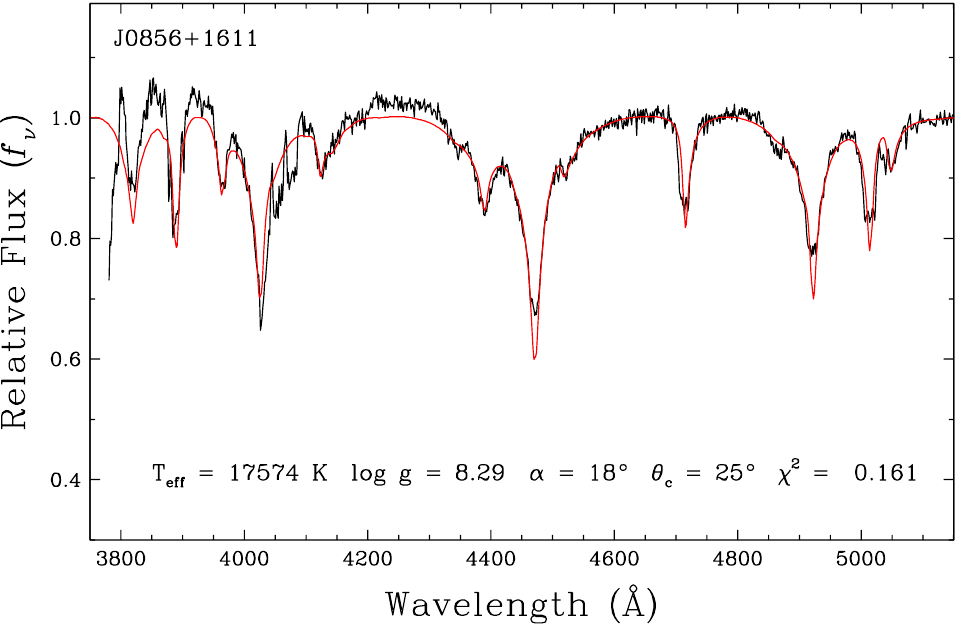}
\includegraphics[width=2.4in, clip=true, trim=1.1in 3.4in 0.9in 3.4in]{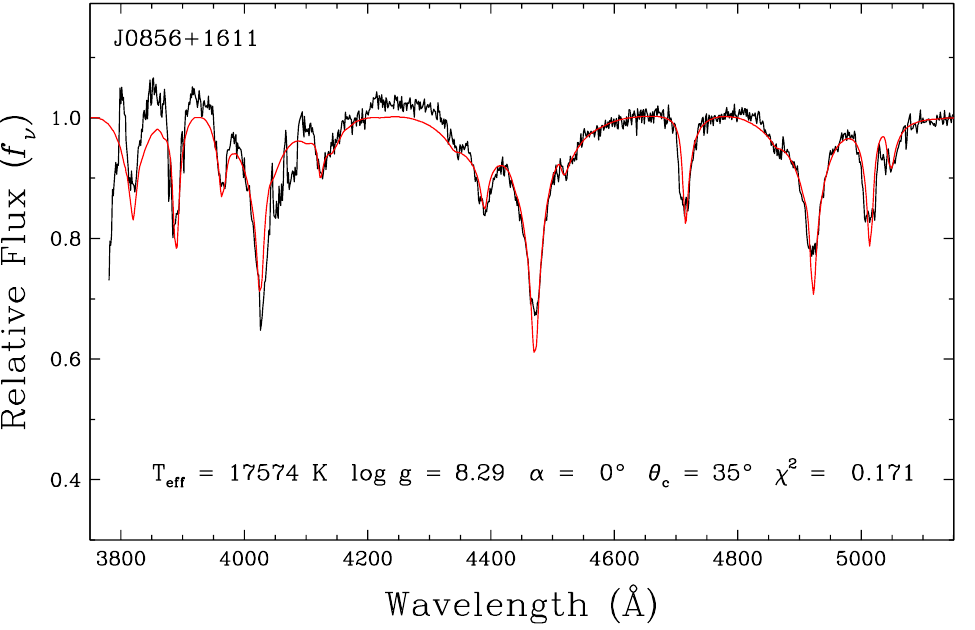}
\caption{Example fits to two J0856+1611 spectra (top and bottom panels) with $\theta_{\rm c}=$ 15, 25, and 35\degree\ (left to right).}
 \label{fig10}
\end{figure*}

Our photometric fit finds $T_{\rm eff} = 17,574$ K, $\log{g}=8.29$, and $M=0.77 \ M_\odot$, again assuming $\log{\rm H/He} = - 3$. Figure \ref{fig10} shows example fits for two of the spectra using inhomogeneous atmosphere models with these parameters. Since the H lines are weak, we are unable to fully constrain the polar cap size, though we rule out a cap size greater than $\approx35$\degree. Instead we show representative fits with $\theta_{\rm c}=$ 15, 25, and $35\degree$. We then use the minimum and maximum $\alpha$ values derived from each of these fixed cap sizes to obtain the best-fit geometry under the assumption of these cap sizes. Table \ref{tab2} shows the derived $\alpha$, $\beta$, and $i$ values assuming these fixed cap sizes.

\begin{table}
\caption{\label{tab2}Geometry values for J0856+1611 assuming fixed cap sizes. All values are in degrees.}
\begin{tabular}{|c|c|c|c|c|} 
 \hline
  $\theta_{\rm c}$ & $\alpha_{\mathrm{min}}$ & $\alpha_{\mathrm{max}}$ & $\beta$ & $i$ \\
 \hline
  15 & 16 & 90 & 37 & 37 \\
  25 & 1 & 37 & 71 & 18 \\
  35 & 0 & 21 & 75 & 15 \\
 \hline
\end{tabular}
\end{table}

\subsection{Non-Variable Objects} 

\begin{figure}
\centering
\includegraphics[width=2.5in,clip=true,trim=0.2in 0.1in 0.4in 0.6in]{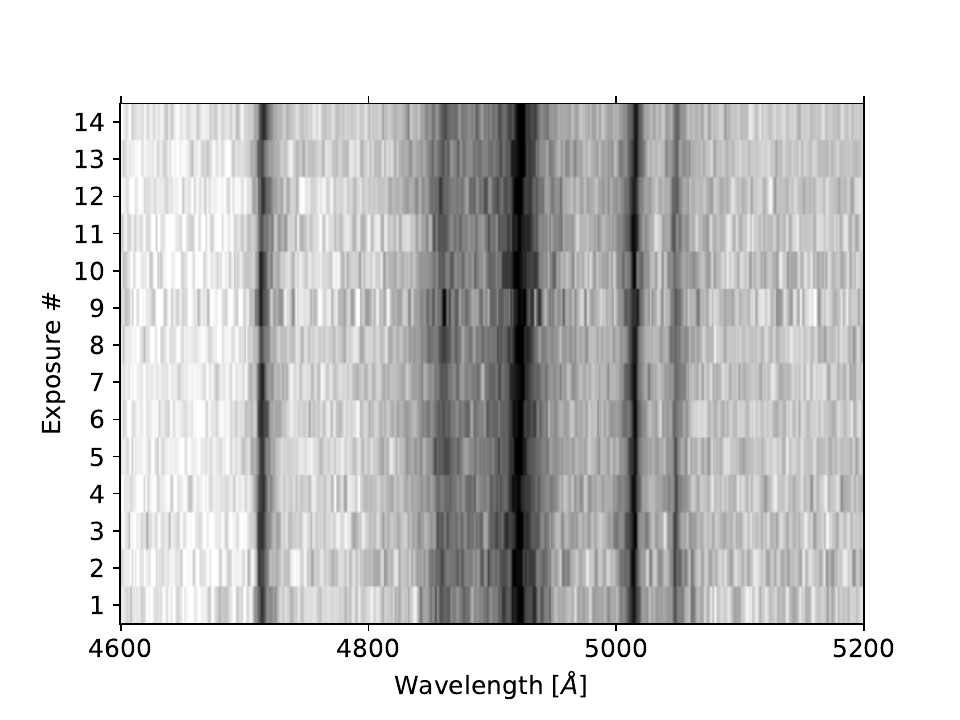}
\includegraphics[width=2.5in,clip=true,trim=1.6in 0.2in 1.7in 0.8in]{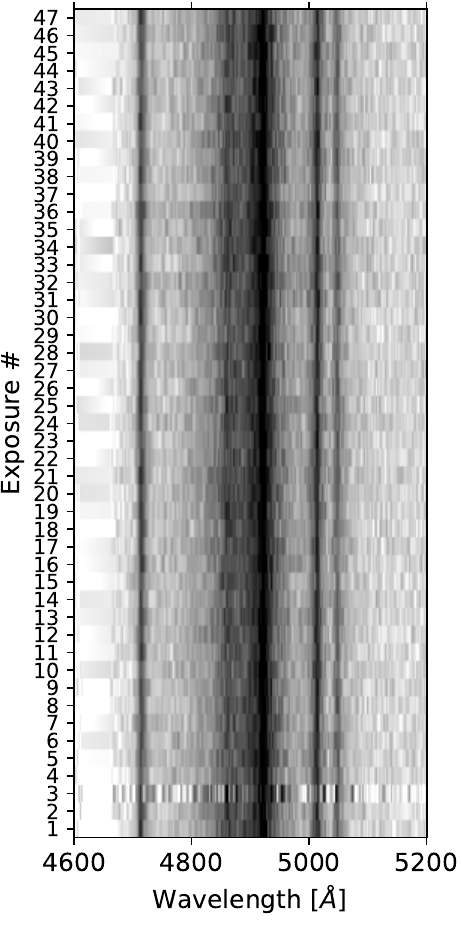}
\caption{Trailed spectra for J0113+3015 (APO, top panel) and J1013+0759 (Gemini, bottom panel).}
\label{fig11}
\end{figure}

Of the seven targets we obtained data for, five do not show spectral variations in our back-to-back exposures. Figure \ref{fig11} shows the trailed spectra for two of these targets. The top panel shows $11\times10$ min exposures for J0113+3015 obtained at APO on UT 20231108 and additional $3\times15$ min exposures from nine days later. The bottom panel shows $47\times3$ min exposures for J1013+0759 obtained at Gemini on UT 2024 January 19. Both hydrogen and helium lines are visible for each star, and there are no significant variations in the strengths of these lines over the course of the $\approx2$ h back-to-back sequences. Figure \ref{figapp} in the Appendix provides the trailed spectra for the remaining three targets, which also do not show any spectroscopic variations. If these white dwarfs have rotation periods longer than a few hours, or if the rotation axis is aligned with the magnetic axis, it may still be possible to hide patchy atmosphere white dwarfs in our data. However, Figure \ref{fig1} shows that all five of these objects are also over-luminous. Hence, they are likely unresolved DA+DB binary systems with periods longer than a few hours. Since \citet{Genest19} obtained excellent fits assuming DA+DB binary systems for these five systems, we do not present any additional modelling of them here.

\section{Discussion}

\subsection{The Class of Double-faced White Dwarfs}

We performed time-series spectroscopy of 7 DBA white dwarfs from \citet{Genest19}, and provided a detailed analysis of two double-faced white dwarfs in this
sample. J0847+4842 is a new discovery; the observed spectroscopic variations can be explained by a double-faced white dwarf where the H-rich magnetic pole comes into view every 6.5 or 8.9 hours. The second target, J0856+1611, was previously identified as a patchy atmosphere white dwarf by \citet{Wesemael01} based on only three spectra available. \citet{Hermes17a} provided a rotation period measurement for J0856+1611 based on K2 photometry, and here we detect spectroscopic variations at the same period. Both J0847+4842 and J0856+1611 are explained well by the oblique rotator model. We now turn our attention to the emerging class of double-faced white dwarfs.

\begin{table*}
\begin{center}
\caption{\label{tab3}Physical parameters of the known double-faced white dwarfs.}
\begin{tabular}{|c|c|c|c|c|c|} 
 \hline
  Object & $T_{\rm eff}$ [K] & Mass [M$_{\odot}]$ & $B_{\rm d}$ [MG] & Rotation Period [hrs] & References\\
 \hline
  J0847+4842 & 14999 & 0.585 & $-$ & 6.5 or 8.9 & This Work \\
  J0856+1611 & 17574 & 0.773 & 1.5 & 5.7 & This Work, \cite{Hermes17a,Hardy23b}\\
  J0910+2105 & 16746 & 0.778 & 0.55 & 7.7 or 11.3 & \cite{Moss24}\\
  ZTF J203349.8+322901.1 & $\sim$35800 & $\sim$1.24 & $-$ & 0.25 & \cite{Caiazzo23}\\
  GD 323 & 26926 & 0.778 & $-$ & 3.5 & \cite{Kilic20}\\
  Feige 7 & 18381 & 1.077 & 39.45 & 2.18 & \cite{Hardy23b,Jewett24}\\
  GALEX J071816.4+373139 & 33942 & 1.27 &$\sim$8 & 0.18 & \cite{Cheng24,Jewett24}\\
 \hline
\end{tabular}
\end{center}
\end{table*}

Table \ref{tab3} summarizes the physical parameters for the 7 known double-faced white dwarfs with patchy atmospheres. These double-faced white dwarfs
are found over a relatively broad temperature range of 15,000 to 36,000 K. More importantly, their mass distribution is skewed. DB white dwarfs in
the solar neighborhood lack the high mass tail seen in DA white dwarfs \citep{Genest19}. For example, the mass distribution of DB white dwarfs
in the 100 pc white dwarf sample and the SDSS footprint is best-explained by a Gaussian with $M=0.586$ and $\sigma=0.036~M_{\odot}$ \citep{Kilic24}.
In addition, there is only a single DB with $M>0.8~M_{\odot}$ in that sample, which is also magnetic. The double-faced white dwarf J0856+1611 happens to be the third most massive DB (among 50 objects) in that sample. \citet{Jewett24} found that there are no normal DBs with $M\geq0.9~M_{\odot}$ in the 100 pc sample
and the Pan-STARRS footprint, those massive DBs are either magnetic or they rotate rapidly. Hence, besides J0847+4842, all other double-faced white dwarfs in Table \ref{tab3} are unusually massive. Massive white dwarfs tend to be magnetic as well, and four of the seven double-faced objects are confirmed to be strongly magnetic. Despite the diversity of $T_{\rm eff}$ in this sample, it seems that magnetism could be what ties the objects together.

\subsection{The Role of Magnetism in Double-Faced White Dwarfs}\label{4.2}

We propose that magnetism is responsible for the emergence of double-faced white dwarfs for two reasons. One is that the frequency of magnetism in this sample is notably higher than that of the DB population. Four out of the seven double-faced targets are confirmed to be magnetic, whereas only 1\% of the DBs in the \citet{Genest19} DB sample are magnetic. 
The second reason is that it is the only plausible way to alter the elemental abundances in such a way that would lead to an inhomogeneous atmosphere. While the float-up and convective dilution/mixing processes greatly alter the surface composition, these impact the entire atmosphere and are expected to lead to well mixed atmospheres. Conversely, the magnetic field should affect any convective processes while still allowing for the flow of material, specifically along the field lines, resulting in surface regions of different abundances. Even in double-faced white dwarfs where Zeeman splitting is not visible in the available low-resolution spectra, magnetism is still invoked to explain patchy atmospheres \citep[e.g.,][]{Caiazzo23}. 

In our paradigm, convection is more inhibited at the poles where the magnetic field strength is higher than at the equatorial regions. This leads to H caps as convective dilution would be unable to convert these regions to He. \citet{Achilleos92} suggested that convection would be inefficient in regions where the field lines are tangent to the surface, thus resulting in He caps and a H ring for Feige 7 in their model. A similar model was used by \citet{Valyavin14} to describe temperature variations in WD 1953$-$011 and suggested convection is less efficient at the equator due to the Lorentz force acting against the direction of gravity. While our H cap model is only a proxy for the possible distribution, \citet{Tremblay15} showed that for a vertical field (i.e. at the poles) the atmosphere becomes radiative for field strengths above $\approx50$ kG. Hence the strength of the local field matters more than the orientation. Since the local field is stronger at the poles, convection is less efficient there, leading to H caps.

Further supporting this hydrogen cap model is the case of J0856+1611. This target shows strong He lines and weak H lines, hence it has a much larger surface area covered by He. If the polar caps are specifically comprised of He in this target, they would need to be quite large as a result. The more likely explanation is a large He equatorial belt and small H caps. It is possible to have He polar caps of similar size to J0847+4842 while having an orientation where we do not see significant variations. If the magnetic axis is not significantly offset from the rotation axis, the visible stellar disk would not significantly change as the object rotates. However, \citet{Wesemael01} made three spectropolarimetric measurements and found varying magnetic field strengths between these observations. In addition, the measured polarizations were of opposite sign from \citet{Putney97}. Hence both magnetic poles are visible at different phases, so the magnetic and rotation axes must be offset. Therefore, the variations we detect are likely due to the small H caps coming in and out of view.

While we propose that magnetism is affecting the convective processes in these white dwarfs, we do not suspect that convection is completely shut down. \citet{Lecavalier17} compared photometric and spectroscopic fits of cool DAs using radiative and convective atmosphere models and found that the photometric and spectroscopic temperatures using convective atmospheres matched while the radiative models did not. They extended this analysis to two weakly magnetic DAs, G217-37 (100 kG, \citealt{Schmidt94}) and LHS 3501 ($\sim$500 kG, \citealt{Maxted00}), and similarly found that the radiative solutions differed, suggesting that convective atmospheres are still prevalent even among magnetic white dwarfs. There is in fact only one particular white dwarf, WD 2105$-$820, in which only radiative atmosphere models are successful \citep{Bedard17,Gentile18}.

The formation of classical DQs (helium-rich atmospheres with traces of carbon) below 10,000 K is attributed to convective dredge-up of carbon from the interior \citep{Pelletier86,BedardDQ}. If magnetism were to fully shut down convection, we would not find any magnetic DQs below 10,000 K, yet there are several confirmed cool magnetic DQs in the 100 pc sample: e.g., GJ 1086, WD 1111+020, and WD 1331+005. GJ 1086 has a field strength of $\sim$10 MG \citep{Ferrario15,Holberg16}. Neither WD 1111+020 nor WD 1331+005 have measured field strengths but they both show distorted carbon bands and circular polarization \citep{Schmidt03}. Hence, the magnetic field does not stop convection fully over the entire surface of the star. However, the strength of convection depends on temperature, and the existence of cool magnetic DQs in principle does not exclude the fact that magnetic fields can successfully inhibit convection at higher temperatures (i.e. WD 2105$-$820).

Interestingly, none of these seven targets have begun core crystallization, hence crystallization induced dynamos cannot explain magnetism in these sources \citep{Isern17,Schreiber21}. \citet{Bagnulo22} found two distinct populations of magnetic white dwarfs: those with weaker fields that grow with time and lower mass ($M < 0.75~M_{\odot}$) which are likely the product of single-star evolution, and those with much stronger fields and higher mass suggesting a merger origin. Two of the double-faced white dwarfs, ZTF J203349.8+322901.1 and J0718+3731, are both ultramassive and have rotation periods of 11-15 min \citep{Caiazzo23,Cheng24}, strongly suggesting a past merger. Feige 7 also has a high mass, is strongly magnetic, and has a rotation period of 2.2 hours \citep{Hernandez24}. Hence, at least three of these double-faced white dwarfs likely originated in binary white dwarf mergers. 

\subsection{Cooler Double-Faced White Dwarfs}

We discussed two mechanisms in section \ref{evol} to convert a DA into a DBA, convective dilution and convective mixing. Convective dilution occurs between $T_{\rm eff}\approx30,000$ and 14,000 K, whereas convective mixing is relevant for temperatures between 14,000 and 6000 K. The seven double-faced white dwarfs presented in Table \ref{tab3} have effective temperatures ranging from about 36,000 down to 15,000 K. Hence, the presence of a magnetic field that affects convective dilution of a DA white dwarf can explain the emergence of this class of double-faced white dwarfs. 

The prediction of this model is that there should also be cooler white dwarfs with inhomogeneous atmospheres. Just like the seven double-faced white dwarfs that likely formed through the magnetic field impacting convective dilution, cooler DAs that normally would go through convective mixing between 14,000 and 6000 K could also end up as patchy atmosphere objects under the presence of a surface magnetic field. Here, we suggest that such objects have already been discovered.

\citet{Rolland15} analyzed 16 magnetic DA white dwarfs with high signal-to-noise ratio optical spectroscopy available and found that 10 of the 16 stars have inconsistent photometric and spectroscopic temperatures. They were able to fit the spectral energy distributions of these magnetic DAs with additional flux from a DC companion, which is needed to explain the relatively shallow H$\alpha$ lines. \citet{Moss23} demonstrated that many of these objects are also rapidly rotating, and they also had to include a DC-offset to match the H line profiles. 

\begin{figure}[hbt!]
\centering
\includegraphics[width=3.5in,clip=true,trim=0in 2.5in 0in 2.4in]{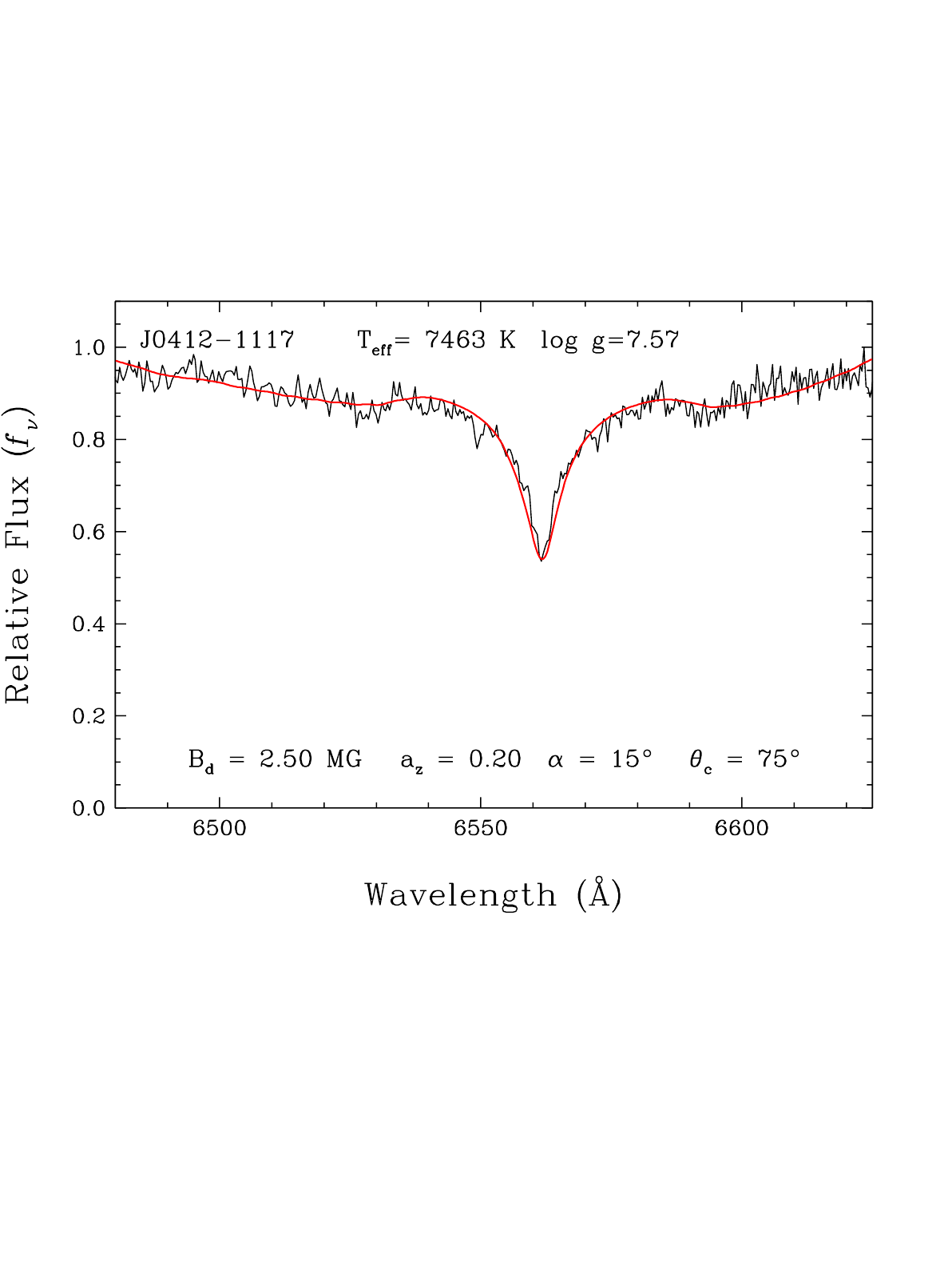}
\includegraphics[width=3.5in,clip=true,trim=0in 2.5in 0.15in 1.8in]{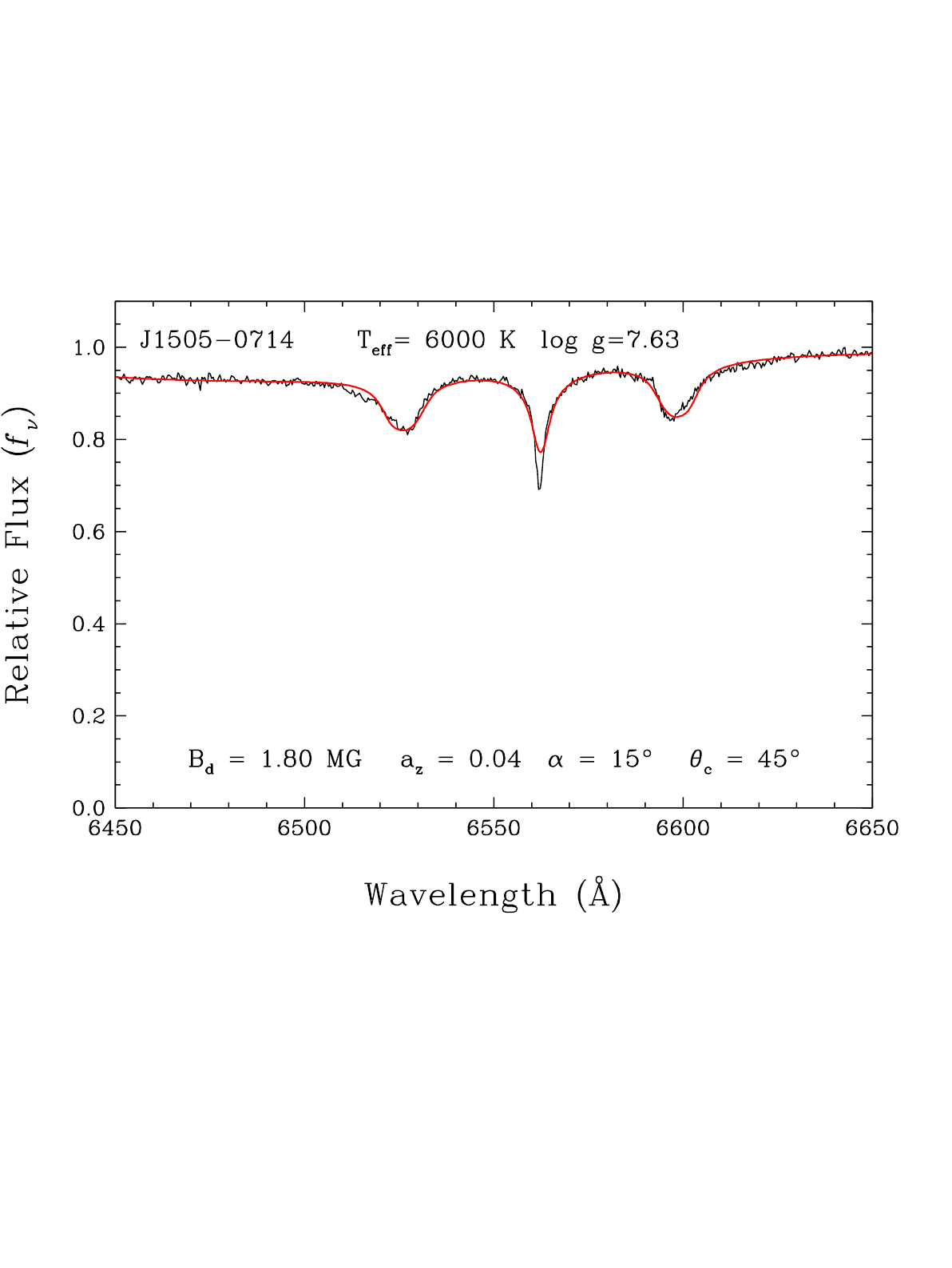}
\caption{Patchy atmosphere fits to the combined spectra of J0412$-$1117 (top) and J1505$-$0714 (bottom). We obtain excellent fits without the need of an unresolved binary companion.}
\label{fig12}
\end{figure}

Figure \ref{fig12} shows our model fits for two of those targets under the assumption of patchy atmospheres: J0412$-$1117 (G160-51) and J1505$-$0714 (GD 175). Here we determine the stellar parameters using the Gaia DR3 distance, Pan-STARRS $grizy$, and 2MASS $JHK$ photometry while assuming $\log{\rm H/He} = - 3$. \citet{Moss23} used the parameters from \citet{Caron23} which assumed pure H atmospheres. Hence we obtain different stellar parameters for G160-51 and GD 175: $T_{\rm eff} = 7463$ K, $\log{g} = 7.57$ and $T_{\rm eff} = 6000$ K, $\log{g} = 7.63$ respectively.

We obtain excellent fits using the polar cap model. The fit for G160-51 is particularly noteworthy, as the strength of the central H$\alpha$ component is neither over nor under-predicted. The obtained cap size of $\theta_c = 75\degree$ is much greater than the other targets we have analyzed, as the observed central line profile is quite broad and deep. These fits are superior to the ones presented in \citet{Moss23} where a DC-offset was required to match the relatively shallow H$\alpha$ line profiles. Hence, these cooler magnetic DAs with shallow H$\alpha$ lines are likely the cooler versions of the double-faced white dwarfs discussed above.

Of the eight targets analyzed by \citet{Moss23}, five show signs of rapid rotation via changes in the line position of the Zeeman-split H$\alpha$ components. The shifts can be interpreted as single stars with their magnetic axis misaligned with the rotation axis, which results in the observer viewing different regions of the complicated magnetic field structure as the object rotates. As we see more of the polar regions, the observed $B_{\rm d}$ is higher, resulting in components that shift further away in wavelength space from the central absorption line. 

\section{Conclusion}

In summary, we present time-series spectroscopy of seven DBA white dwarfs, six of which were previously identified as unresolved DA+DB binary candidates. One of these candidates, J0847+4842, shows extreme variations in the absorption lines over the span of our exposure sequences. Hence this unresolved binary candidate is clearly a single white dwarf with an inhomogeneous atmosphere, which results in varying line strength as the object rotates. We constrained the rotation period to either 6.5 or 8.9 hours, and successfully implemented a polar cap model to obtain fits to our spectra. We fix the cap size to 40\degree\ and allow the angle between the magnetic axis and plane of the sky to vary in our fits, which lets us constrain the system geometry using the oblique rotator model. We determine that $\beta = 33\degree$ and $i = 33\degree$, so the angles between the magnetic axis/line-of-sight and the rotation axis are equal. Hence at certain phases, we are looking directly at the polar cap, which leads to a spectrum that looks effectively like a DA spectrum. As the target rotates, we see more of the equatorial region, which leads to the appearance of He lines. 

A separate target we observed is a known variable DBAH, J0856+1611, with a rotation period of 5.7 hours. We detected variations in the strength of the weak H lines, and obtained the same previously determined rotation period. While we do obtain good fits with our polar cap model, the H lines are too weak to put a tight constraint on the cap size. Regardless, it is clear this target has a patchy atmosphere, with likely small caps and an orientation where we mostly view the equatorial region at all phases.  

Given the several patchy-atmosphere objects known at this time, we are able to define the class of double-faced objects for the first time. While most objects maintain a homogeneous atmosphere throughout their evolution, this class likely forms as a result of the magnetic field influencing the motion and mixing of either H or He, creating an atmosphere with varying surface abundances. Since magnetism is quite rare in He-dominated white dwarfs, it follows that these inhomogeneous atmospheres are quite rare among the DBA population. 

The targets discussed in this work are at effective temperatures where convective dilution is expected to either have already occurred or is currently ongoing. It then follows that double-faced systems could form at cooler temperatures where convective mixing should occur. We identify such a population among the cool magnetic DA white dwarf sample, where the H$\alpha$ line is shallower than expected based on pure hydrogen atmosphere models. 

An important step in analyzing this class is further modelling on how the magnetic field influences convection. If modelling the transport mechanisms can indeed reproduce these atmospheres, that will further solidify the idea that magnetism is the driving force behind the creation of these objects, even if magnetism is not detected in the observations. Of course this also means better modelling of complex fields in general is needed. In our framework, the dipole field is stronger at the polar regions compared to the equatorial regions, with inefficient mixing at the poles. This framework fits well with the traditional dipole field that is used in modelling magnetic white dwarfs. However, the true field structure in these targets could be more complicated, particularly in the case of ZTF J203349.8+322901.1 since it has two opposing faces \citep{Caiazzo23}. Nevertheless, while the exact nature of the atmosphere geometry cannot be directly confirmed due to the complexities in how magnetism affects convection, we are confident that magnetism is the source of the inhomogeneities in this class of objects. \\

We are grateful to Antoine Bédard for useful discussions.
We thank the students of the Spring 2024 Advanced Observatory Methods class at OU for obtaining spectra of J0847+4842 and J0856+1611 during their telescope training on UT 2024 April 4 and April 5.
This work is supported in part by the NSF under grant AST-2205736, the NASA under grants 80NSSC22K0479, 80NSSC24K0380, and 80NSSC24K0436, the NSERC Canada, the Fund FRQ$-$NT (Québec), and the Canadian Institute for Theoretical Astrophysics (CITA) National Fellowship Program.

Based on observations obtained with the Apache Point Observatory 3.5-meter telescope, which is owned and operated by the Astrophysical Research Consortium. 

Observations reported here were obtained at the MMT Observatory, a joint facility of the Smithsonian Institution and the University of Arizona. 

Based on observations obtained at the international Gemini Observatory, a program of NSF’s NOIRLab, which is managed by the Association of Universities for Research in Astronomy (AURA) under a cooperative agreement with the National Science Foundation on behalf of the Gemini Observatory partnership: the National Science Foundation (United States), National Research Council (Canada), Agencia Nacional de Investigación y Desarrollo (Chile), Ministerio de Ciencia, Tecnología e Innovación (Argentina), Ministério da Ciência, Tecnologia, Inovações e Comunicações (Brazil), and Korea Astronomy and Space Science Institute (Republic of Korea).

\bibliographystyle{aasjournal}
\bibliography{MossClass.bib}

\appendix

\begin{figure}[hbt!]
\includegraphics[width=2.3in,clip=true,trim=0.2in 0.8in 0.4in 1.4in]{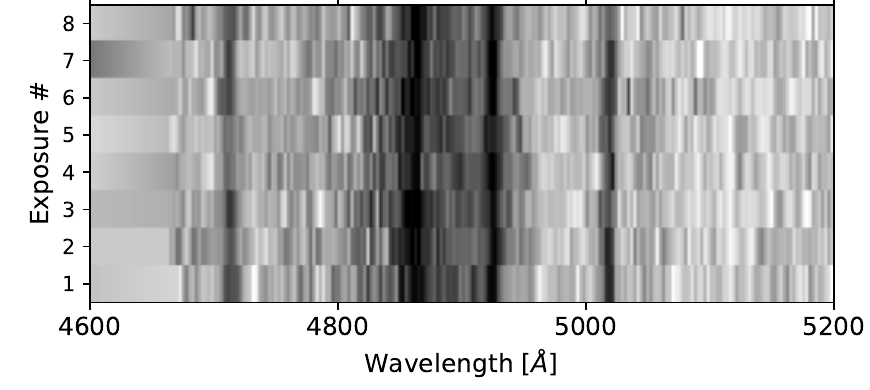}
\includegraphics[width=2.3in,clip=true,trim=1.1in 0in 1.2in 0.4in]{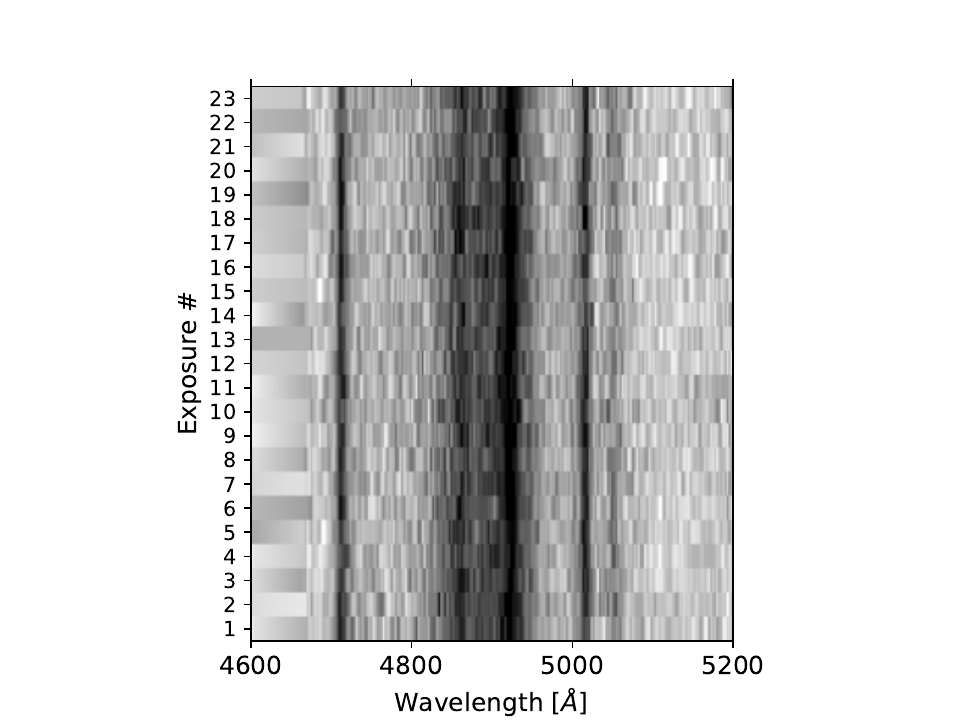}
\includegraphics[width=2.3in,clip=true,trim=1.2in 0.2in 1.4in 0.8in]{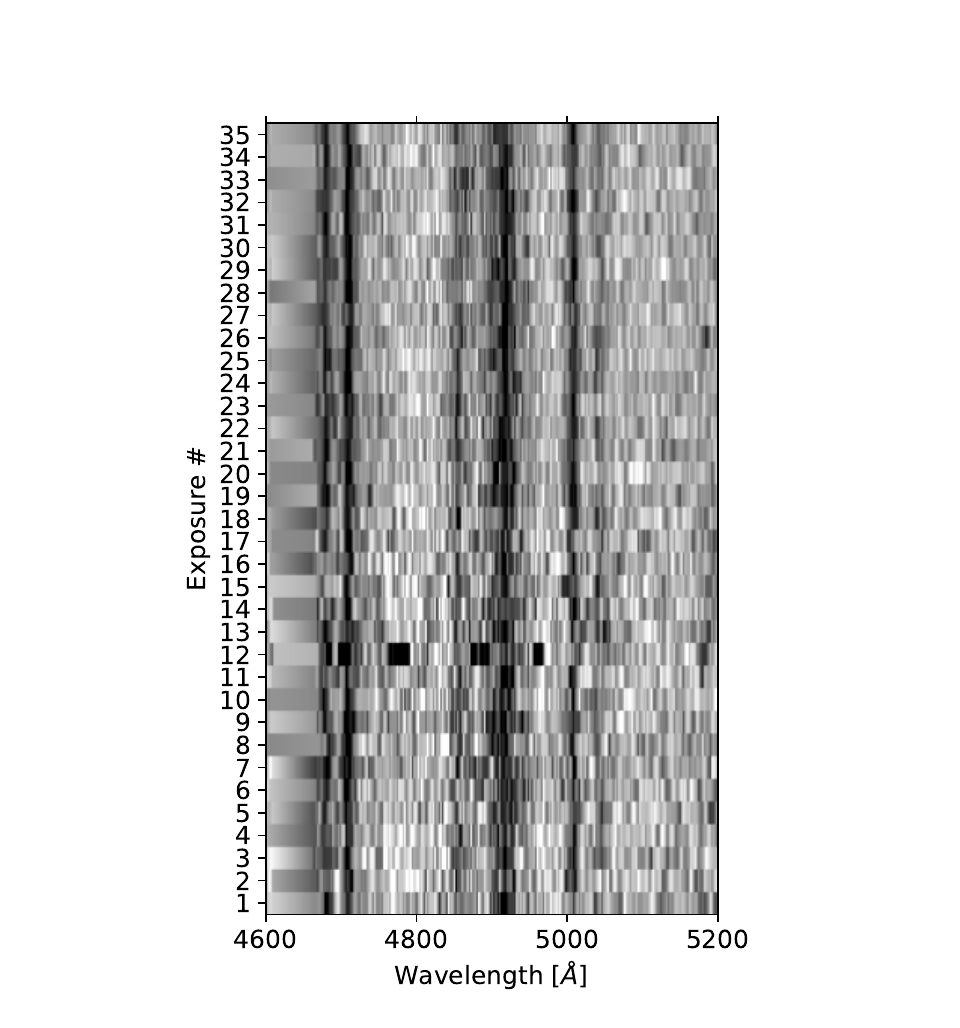}
\caption{Trailed Gemini spectra for our remaining targets that do not show spectral variations: J1127+3525, J1136+3204, and J1406+5627 (from left to right).}
\label{figapp}
\end{figure}

\end{document}